\documentclass[12pt]{article}
\DeclareUnicodeCharacter{2032}{'}
\usepackage{graphicx}
\usepackage{natbib} 
\usepackage{url} 
\usepackage{authblk}
\newcommand{\norm}[1]{\left\lVert#1\right\rVert}

\newcommand{\blind}{0}

\usepackage{amssymb}
\usepackage{bm}
\usepackage{amsmath}
\usepackage{amsthm}
\newtheorem{proposition}{Proposition}

\newtheoremstyle{exampleupright}
  {\topsep}                    
  {\topsep}                    
  {\normalfont}               
  {}                          
  {\itshape}                  
  {.}                         
  { }                         
  {\thmname{#1}\ \thmnumber{#2}\thmnote{ (#3)}} 

\usepackage{subcaption}
\theoremstyle{exampleupright}

\usepackage{booktabs}
\usepackage{multirow}
\usepackage{appendix}
\usepackage{rotating}
\usepackage{mathtools}
\usepackage{graphicx}
\usepackage{algorithm}
\usepackage{algpseudocode}
\usepackage{makecell}
\usepackage{changes} 
\setdeletedmarkup{\color{red}\sout{#1}}
\usepackage{placeins}
\usepackage{hyperref}

\DeclareMathOperator*{\argmax}{\arg\!\max}

\begin{document}

\def\spacingset#1{\renewcommand{\baselinestretch}%
{#1}\small\normalsize} \spacingset{1}

\if0\blind
{
  \title{\bf Space-Filling One-Factor-At-A-Time Designs}
  \author{Wei-Yang Yu}
  \author{V. Roshan Joseph\thanks{Corresponding author: roshan@gatech.edu}}
  \affil{H. Milton Stewart School of Industrial and Systems Engineering, \\
        Georgia Institute of Technology, Atlanta, GA 30332}
  \maketitle
} \fi

\if1\blind
{
  \bigskip
  \bigskip
  \bigskip
  \begin{center}
    {\LARGE\bf Space-Filling One-Factor-At-A-Time Designs}
\end{center}
  \medskip
} \fi

\bigskip
\begin{abstract}
Space-filling designs are commonly used in deterministic computer experiments. However, they cannot do factor screening without any modeling assumptions, which makes them inefficient when only a small subset of input factors is important. Recently developed screening designs, such as MOFAT designs, can identify important factors without a modeling step but lack space-filling properties, limiting their usefulness for surrogate modeling. In this article, we propose a new class of screening designs that improves the space-fillingness while retaining their screening capability. Through several numerical examples, we demonstrate that the proposed designs offer clear advantages over existing designs.

\end{abstract}

\spacingset{1.7} 
\section{Introduction}
Computer models are essential tools for simulating complex systems, such as for optimizing materials crystal structures using density functional theory \citep{krishna2023adaptive}, predicting damage of Aircraft-UAV collisions using finite element models \citep{liu2023statistical}, or improving inverse scattering using multi-fidelity simulations \citep{sung2025advancing}. These models are typically deterministic, meaning that the same input always produces the same output. To realistically represent the physical systems, computer models have become increasingly complex, making each model evaluation computationally intensive. Therefore, experimental designs are necessary to efficiently explore the input space. Surrogate models, such as Gaussian processes (GPs, \citet{gramacy2020surrogates}), are then employed to approximate the expensive computer model, making optimization and exploration more efficient in future investigations.


Space-filling designs are widely used as experimental designs in deterministic computer experiments \citep[Ch. 4]{joseph2026experimental}. These designs aim at placing points everywhere in the input space with as few gaps as possible, leading to robust performance for learning an unknown response surface. Since effect sparsity is common in almost all systems---meaning that only a small subset of input factors is influential to the output---considerable effort has been devoted to improving the projections of space-filling designs onto lower-dimensional subspaces. To improve the one-dimensional projections, \citet{morris1995exploratory} proposed using a \textit{maximin Latin hypercube design} (MmLHD) that maximizes the minimum distance among the points in a Latin hypercube design (LHD). \citet{joseph2015maximum} proposed the \textit{maximum projection design} (MaxPro) that ensures good projections to all subspaces of the input factors. However, improving projection properties alone is not the most effective strategy. By identifying the important factors at an early stage, we can focus modeling efforts on the smaller active subspace, which can help us develop a more efficient and accurate surrogate model. This task is known as ``factor screening" in sensitivity analysis \citep{saltelli2008global} and ``variable selection" in statistics \citep{miller2002subset}. 

Traditionally, screening and modeling are done sequentially \citep{myers2016response}.
However, this sequential strategy can become infeasible for expensive computer experiments where all runs must be launched in a single batch to exploit parallel computing. Therefore, a one-shot design that supports both factor screening and surrogate modeling can be beneficial. Moreover, such designs can also make the sequential experiments more efficient when used as initial designs \citep{song2026efficient}. However, because space-filling designs vary all factors simultaneously, individual input effects cannot be directly isolated. Factor screening on space-filling designs therefore requires either fitting a surrogate model \citep{oakley2004probabilistic} or using a sensitivity measure that involves additional assumptions \citep{marrelICSCREAMMethodologyIdentification2022}, which can be inefficient.

\begin{figure}[h!]
    \centering
    \includegraphics[scale=0.069]{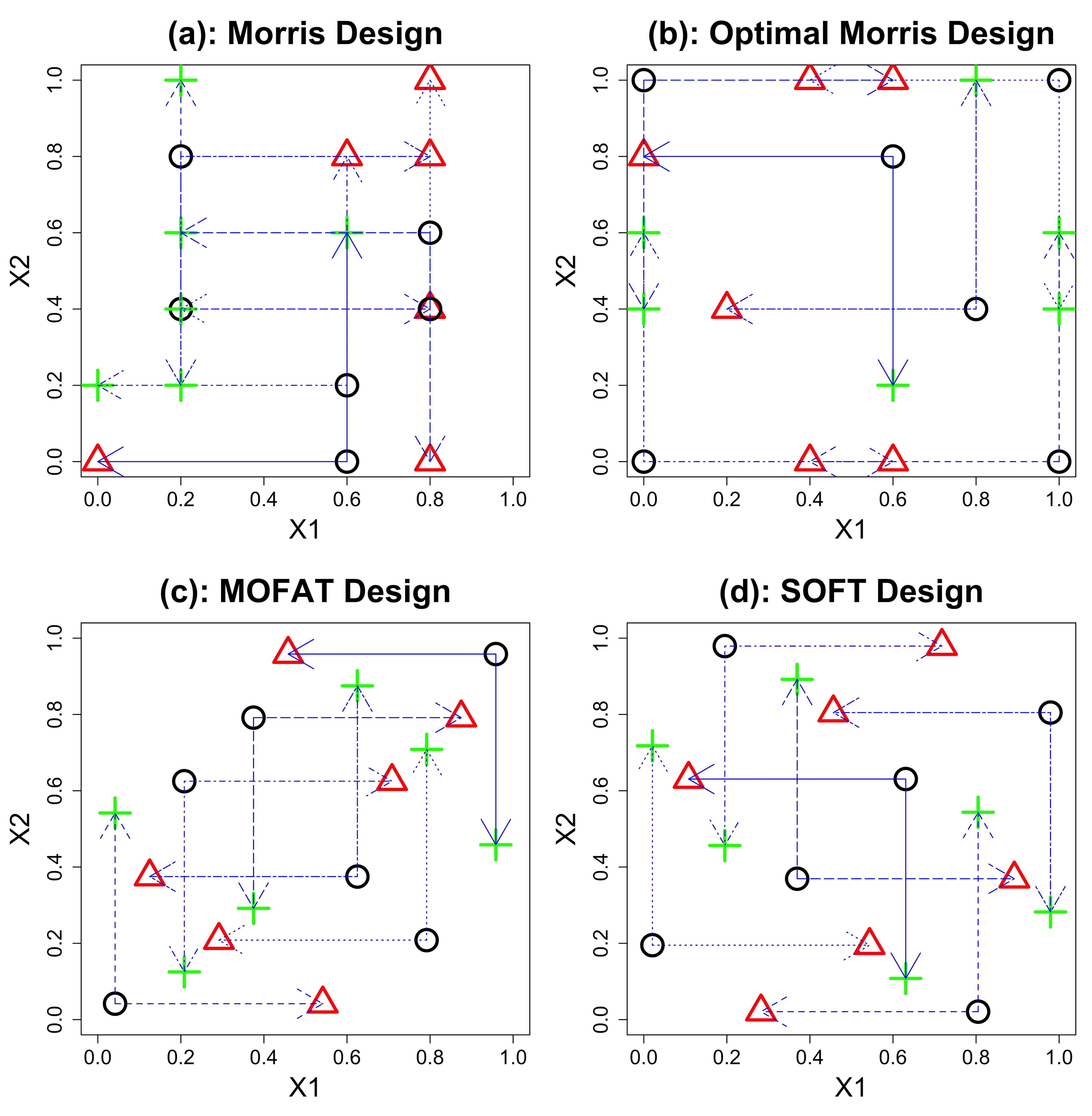}
    \caption{Illustration of four OFAT-type designs: (a) a realization of Morris design, (b) optimal Morris design using the method of \cite{campolongo2007effective}, (c) MOFAT design, and (d) the proposed SOFT design.}
    \label{fig:previous OFATs}
\end{figure}

In contrast, changing one factor at a time is the simplest and most intuitive way to perform factor screening without fitting a surrogate model or imposing additional assumptions. However, a single \textit{one-factor-at-a-time} (OFAT) design can only estimate local sensitivity and fails to explore the entire input space or detect interactions \citep{czitrom1999one,saltelli2010avoid}. To address these limitations, \citet{morris1991factorial} proposed randomly placing a collection of OFAT designs in the input space. Each OFAT provides an estimate of local sensitivity, which are aggregated to obtain the factor's global importance. However, the random placement of OFATs can lead to poor coverage of the input space, as shown in Figure~\ref{fig:previous OFATs}(a). To improve the coverage, \citet{campolongo2007effective} proposed to maximize the sum of pairwise distances among the OFATs' design points. An illustration is shown in Figure~\ref{fig:previous OFATs}(b). Although this strategy encourages OFATs to be placed as far apart as possible, it tends to push points toward the boundary of the input space, leaving large regions of the interior unexplored. Instead of fixating on OFAT designs, \citet{pujol2009simplex} proposed using a collection of simplex designs to improve the projection properties. More generally, \citet{santiagoImprovedSensitivityMorris2012} proposed to select a collection of constellations from an existing space-filling design. A first-order polynomial is then fitted within each of these local designs (simplex or constellation) to estimate the local sensitivity, but they can introduce bias in the estimation of global sensitivity indices when dealing with highly complex nonlinear models.

There are many Monte Carlo methods for estimating the global sensitivity indices \citep{saltelli2008global, da2021basics} of which Sobol' design \citep{sobol2001global} is a popular choice. \citet{xiao2023maximum} recently pointed out the similarities between the Sobol' design and Morris screening design. See also \cite{saltelli2010variance}. Exploiting their common OFAT structure,  \citet{xiao2023maximum} proposed an improved version called the \textit{maximum one-factor-at-a-time} (MOFAT) design. An illustration of the MOFAT design is provided in Figure~\ref{fig:previous OFATs}(c). Clearly, it is an improvement to the optimal Morris screening design, but it still misses some areas in the input space. Therefore, in this article, we propose a new class of OFAT designs, called \textit{Space-filling One-Factor-at-a-Time} (SOFT) designs, that achieves optimal space-filling properties while retaining their screening capability. The SOFT design is shown in Figure~\ref{fig:previous OFATs}(d). We can see that it provides not only excellent coverage of the input space but also good screening ability due to its OFAT structure.

Although SOFT designs improve space-fillingness within the class of OFAT designs, the OFAT structure itself can limit how effectively the experimental budget is used for surrogate modeling on a smaller active subspace. To alleviate this limitation, we further develop Augmented SOFT (A-SOFT) designs, which allocate part of the budget to space-filling points that are unrestricted by the OFAT structure. This improves the flexibility for exploring the input space at the cost of fewer OFAT comparisons and, consequently, some loss in factor-screening accuracy. The augmented design therefore provides a trade-off between factor screening and space-fillingness, with the goal of achieving more robust surrogate-modeling performance.

The article is organized as follows. In Section~\ref{sec:Background: MOFAT Designs}, we first review the existing OFAT designs and introduce the Gaussian process modeling with the multiplicative inverse multiquadric (MIM) kernel. We will then develop our proposed SOFT design in Section~\ref{sec:SOFT Designs}. Several numerical studies are provided in Section~\ref{sec:Numerical Studies} to demonstrate the advantages of the proposed method. Motivated by these results, Section~\ref{sec:Augmented SOFT Designs} introduces A-SOFT designs. We conclude the article with some remarks in Section~\ref{sec:Conclusion}.



\section{Background}
\label{sec:Background: MOFAT Designs}
Suppose that there are $p$ input factors $\mathbf{x} = (x_1, \dots, x_p)'$ and a scalar output $y$. Without loss of generality, we define the input space $\mathcal{X}$ as $[0,1]^p$. 

\subsection{Morris Designs}
\citet{morris1991factorial} defined elementary effects to estimate the local sensitivity of factors by changing one factor at a time: 
\[
e_{i}(\mathbf{x}) = \frac{y(\mathbf{x}^{(i)}(\Delta)) - y(\mathbf{x})}{ \Delta}, i = 1, \dots, p,
\]
where $\mathbf{x}^{(i)}(\Delta) = (x_1, \dots, x_i \pm \Delta, \dots, x_p)$ and the $+$ or $-$ sign for $\Delta$ is chosen such that $x_i + \Delta \in [0,1]$ or $x_i - \Delta \in [0,1]$. By randomly choosing $l$ base runs $\{\mathbf{x}_j\}_{j=1}^l$ from an $L$-level regular grid: $\{0, 1/(L-1), \dots, 1\}^p$ and setting $\Delta$ as a multiple of $1/(L-1)$, the Morris design is a collection of $l$ OFAT designs $\bm O^1,\ldots,\bm O^l$, defined as
\begin{equation}
\label{eq:OFAT design}
\begin{aligned}
\bm{D}
&=
\begin{pmatrix}
\bm{O}^1 \\
\bm{O}^2 \\
\vdots \\
\bm{O}^l
\end{pmatrix},
\; \text{where }
\bm{O}^j
=
\begin{pmatrix}
\mathbf{x}_j\\
\mathbf{x}_j^{(1)}\\
\vdots \\
\mathbf{x}_j^{(p)}
\end{pmatrix}, \; j = 1, \dots, l.
\end{aligned}
\end{equation}
Let $e_{i(j)}$ be the elementary effect of the $i$th factor estimated using the $j$th OFAT, $j = 1, \dots, l$. Then, the global sensitivity can be summarized by the following measures:
\[
\mu_i = \frac{1}{l} \sum_{j=1}^l e_{i(j)} \text{ and } \sigma_i = \sqrt{\frac{1}{l-1} \sum_{j=1}^l (e_{i(j)} - \mu_i)^2},
\]
which can be used to screen  unimportant factors.

An illustration of the Morris design with $p = 2$ and $l = 6$ is shown in Figure~\ref{fig:previous OFATs}(a). The base runs $\{\mathbf{x}_j\}_{j=1}^l$ are represented by black circles, while $\{\mathbf{x}_j^{(1)}\}_{j=1}^l$ and $\{\mathbf{x}_j^{(2)}\}_{j=1}^l$ are shown as red triangles and green crosses, respectively. This illustration uses the \textit{standard} structure according to the classification by \citet{daniel1973one} (also called as radial designs in \cite{saltelli2010variance}), where all changes are based on the same run. Instead of the standard structure, one may use the \textit{strict} structure, where a run is based on the previous run. As seen in the figure, the design does not provide good coverage of the input space, and duplicated points are present. \citet{campolongo2007effective} proposed a strategy to optimize the coverage of the Morris design. Their idea is to first generate a large number of OFAT designs and then choose $l$ of them that are as far apart as possible. The ``distance" between two OFATs $\bm{O}^m$ and $\bm{O}^k$ is defined as
\[
d_{mk} =
\begin{cases}
\displaystyle
\sum_{i=1}^{p+1} \sum_{j=1}^{p+1}
\norm{\bm{O}^m_i - \bm{O}^k_j}
,
& \text{for } m \neq k, \\[2ex]
0, & \text{otherwise},
\end{cases}
\]
where $\bm{O}^m_i$ and $\bm{O}^k_j$ denote the $i$th and $j$th rows of $\bm{O}^m$ and $\bm{O}^k$, respectively.
The optimal $l$ OFATs are then selected by maximizing 
\begin{equation}
    \label{eq:campolongo}
    \sqrt{\sum_{m=1}^{l-1} \sum_{k = m+1}^l d_{mk}^2}.
\end{equation}
Figure~\ref{fig:previous OFATs}(b) shows the optimal Morris design. By construction, the criterion \eqref{eq:campolongo} discourages duplicated points. However, it does not improve the design projections. Moreover, this strategy tends to push points toward the boundary, leaving large unexplored regions in the interior. Implementations of the Morris design and the optimal Morris design are available in the R package \texttt{sensitivity} \citep{Iooss2026sensitivity}.

\subsection{MOFAT Designs}
\begin{figure}[h!]
    \centering
    \includegraphics[scale=0.12]{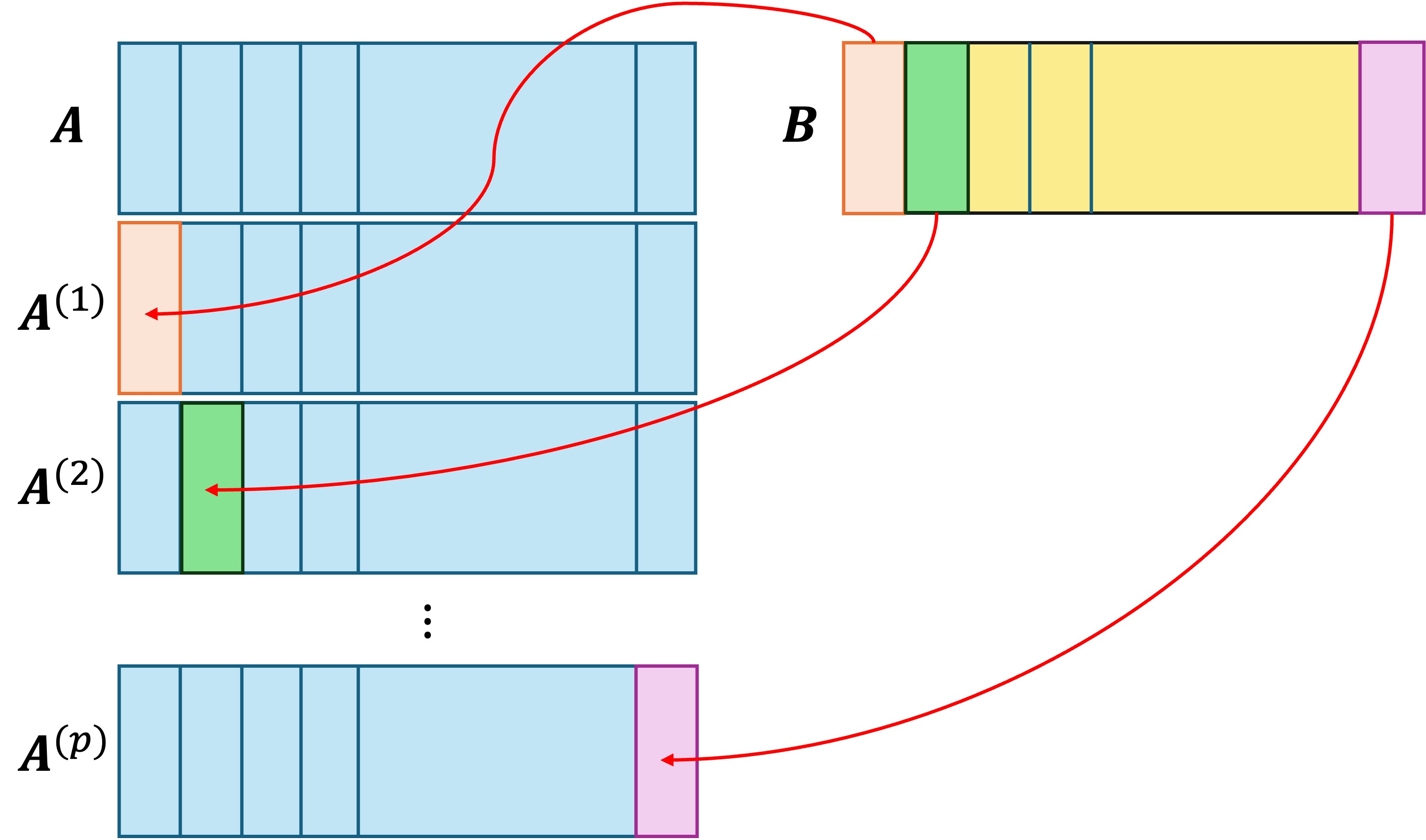}
    \caption{Monte Carlo-based method proposed by \citet{sobol2001global}.}
    \label{fig:sobol}
\end{figure}

Assuming that the underlying function between $y$ and $\mathbf{x}$ is square integrable, \citet{Sobol_1990aa, Sobol_1993bb} shows that the variance of $y$ can be decomposed into the variance resulting from the main effects, two-factor interactions, etc., as follows:
\[
\text{var}(y) = V = \sum_{i=1}^p V_i + \sum_{i<j}^p V_{ij} + \dots + V_{12\dots p},
\]
where the variance components are $V_i = \text{var}_{x_i}\{E_{\mathbf{x}_{\sim i}}(y|x_i)\}$, $V_{ij} = \text{var}_{x_i,x_j} \{E_{\mathbf{x}_{\sim ij}}(y|x_i,x_j)\} - V_i - V_j$, etc., and $\mathbf{x}_{\sim i}$ represents the set of all factors except $x_i$.
Two popular variance-based sensitivity measures derived from this decomposition are the first-order Sobol' index \citep{Sobol_1993bb} and the total Sobol' index \citep{homma1996importance}. The former focuses on factors' main effects, while the latter considers the overall contribution including main effects and interactions. Since the goal of factor screening is to assess the overall importance of each factor to the output, we focus on the total Sobol' index, defined as
\begin{equation*}
    t_i = \frac{V_i^{tot}}{\text{var}(y)},
    \text{ where } V_i^{tot} = \sum_{I\ni i} V_I= E_{\mathbf{x}_{\sim i}} \{\text{var}_{x_i}(y|\mathbf{x}_{\sim i}) \}, \; i = 1, \dots, p.
\end{equation*}
To estimate $t_i$, \citet{sobol2001global} proposed a Monte Carlo-based method that starts by randomly generating two $l \times p$ matrices, $\bm{A}$ and $\bm{B}$. For each $i = 1, \dots, p$, a matrix $\bm{A}^{(i)}$ is constructed by replacing the $i$th column of $\bm{A}$ with the $i$th column of $\bm{B}$. An illustration of the method is provided in Figure~\ref{fig:sobol}.

Sobol' design is then defined as 
\begin{equation}
\label{eq:sobol' design}
\bm{D} =
\begin{pmatrix}
\bm{A} \\
\bm{A}^{(1)} \\
\vdots \\
\bm{A}^{(p)}
\end{pmatrix}.
\end{equation}
A commonly used estimator for $t_i$ is the Jansen-Sobol' (J-S) estimator \citep{jansen1999analysis}:
\begin{equation}
\label{eq:ti}
\hat{t}_i = \frac{ \hat{V}^{tot}_i}{\widehat{\text{var}}(y)},
\text{ where } \hat{V}^{tot}_i = \frac{1}{2l} \sum_{j=1}^l \{ y(\bm{A}_j) - y(\bm{A}_j^{(i)}) \}^2.
\end{equation} 
Here, $\bm{A}_j$ and $\bm{A}_j^{(i)}$ denote the $j$-th row of $\bm{A}$ and $\bm{A}^{(i)}$, respectively. The sample variance $\widehat{\text{var}}(y)$ is given by $\sum_{j=1}^n (y_j - \bar{y})^2 / (n-1)$, where $\bar{y}=\sum_{j=1}^n y_j/n$ is the sample mean and $n = l\times(p+1)$ is the run size. If $\hat{t}_i$ is greater than $\hat{t}_j$ $(i,j = 1, \dots, p)$, we can conclude that factor $i$ is more important than factor $j$. Thus, $\{\hat{t}_i\}_{i=1}^p$ can be used to select the $k$ most important factors $(k \leq p)$ or to screen out factors with $\hat{t}_i \approx 0$.

Based on the same structure, \citet{xiao2023maximum} proposed an improvement of the Sobol' design, which is called the maximum one-factor-at-a-time (MOFAT) design. In their construction, $\bm{A}$ and $\bm{B}$ are two Latin hypercube designs (LHDs, \citet{mckay1979comparison}) with equally spaced levels given by $\{0.5/l, 1.5/l, \dots, (l-0.5)/l\}$. To maximize the capacity of identifying important factors, $\bm{A}$ and $\bm{B}$ are chosen to maximize the expected total Sobol' indices under the assumption that the responses are generated from a Brownian process \citep{zhang2014fractional}. Thus, the MOFAT design is defined as $\bm{D}$ in \eqref{eq:sobol' design} that maximizes 
\begin{equation}
    \label{eq:MOFAT criterion}
    \sum_{j=1}^l |\bm{A}_{ji} - \bm{A}_{ji}^{(i)}| \text{ for all } i = 1. \dots, p,
\end{equation}
where $\bm{A}, \bm{A}^{(i)}$ $(i = 1, \dots, p)$ are all LHDs. 
The optimization can be simplified by introducing the following transformation: 
\begin{equation}
    \label{eq:mofat transformation}
    T_l(x) = x- \frac{\lfloor l/2 \rfloor}{l} + \mathbb{I}(x < 0.5), \; x \in \{0.5/l, 1.5/l, \dots, (l-0.5)/l\},
\end{equation}
where $\lfloor x \rfloor$ is the largest integer not exceeding $x$ and $\mathbb{I}(\cdot)$ denotes the indicator function. Now, given an arbitrary LHD $\bm{A}$, the matrix $\bm{B}$ is obtained by $\bm{B}_{ij} = T_l(\bm{A}_{ij})$. \citet{xiao2023maximum} showed that the design $\bm{D}$ constructed using these two matrices $\bm{A}$ and $\bm{B}$ will be a MOFAT design for \textit{any LHD}  $\bm A$. 

To further improve the projection properties of the MOFAT design, the values in $\bm{A}$ and $\bm{B}$ can be adjusted as follows:
\begin{equation}
\label{eq:update_minus}
x \leftarrow 
\begin{cases}
x - \dfrac{0.25}{l}, & \text{if } x < 0.5, \\
x + \dfrac{0.25}{l}, & \text{if } x \geq 0.5,
\end{cases}
\end{equation}
\begin{equation}
\label{eq:update_plus}
x \leftarrow 
\begin{cases}
x + \dfrac{0.25}{l}, & \text{if } x < 0.5, \\
x - \dfrac{0.25}{l}, & \text{if } x \geq 0.5.
\end{cases}
\end{equation}
If \eqref{eq:update_minus} is used for $\bm{A}$, then \eqref{eq:update_plus} should be used for $\bm{B}$, and vice versa. This adjustment doubles the number of levels for each factor (from $l$ to $2l$). An illustration of a MOFAT design with $p = 2$ and $l = 6$ is provided in Figure~\ref{fig:previous OFATs}(c), where $\bm{A}$, $\bm{A}^{(1)}$, and $\bm{A}^{(2)}$ are represented by black circles, red triangles, and green crosses, respectively. Although better than the Morris designs shown in Figure~\ref{fig:previous OFATs}(a-b) , the design is still not space-filling as there are large unexplored regions in the top-left and bottom-right corners.

\subsection{Gaussian Process Modeling with OFAT Designs}
\label{subsec:Gaussian Process Modeling}
Data from experiments are usually analyzed using a Gaussian process (GP) \citep{santner2003design, gramacy2020surrogates}. However, fitting a GP model to a high-dimensional dataset is not an easy task. \citet{song2026efficient} recently showed that if the experimental design is an OFAT design, then the fitting task can be greatly simplified.


Consider the GP model 
\begin{equation*}
    y(\mathbf{x}) \sim \text{GP}(\mu, \sigma^2 R(\cdot)),
\end{equation*}
where $R(\cdot)$ is the correlation function. To simplify the fitting of the GP model, \citet{song2026efficient} proposed the following correlation function called multiplicative inverse multiquadric (MIM) kernel: 
\begin{equation}
    \label{eq:MIM kernel}
    R(\mathbf{x}, \mathbf{x}') = \prod_{i=1}^p \left(1 + \frac{(x_i-x_i')^2}{\theta^2_i}\right)^{-\alpha_i}, \; \alpha_1, \dots, \alpha_p > 0.
\end{equation}
Assume that the changes of each factor within an OFAT are constant $\Delta$. This is true for all Morris designs by construction. For a MOFAT design, this holds when $l$ is even. Under this condition, we have the relationship
\[
\frac{\mathbb{E}\!\left[\widehat{V}_i^{\mathrm{tot}}\right]}
{\operatorname{var}(y)}
=
\frac{1}{\beta}
\left\{
1-
\left(
1+\frac{\Delta^2}{\theta_i^2}
\right)^{-\alpha_i}
\right\},
\]
where $\hat{V}_i^{tot}$ is the numerator of $\hat{t}_i$ and $\beta = \operatorname{var}(y) / \sigma^2$ is a tuning parameter. By replacing the left-hand side with the estimated total Sobol’ index and plugging in estimates of \(\beta\) and \(\theta_i\), we obtain the following estimate of \(\alpha_i\):
\begin{equation}
\label{eq:analytically solve alpha}
\hat{t}_i \approx 
\frac{1}{\beta}
\left\{
1 - \left(1 + \frac{\Delta^2}{\theta_i^2}\right)^{-\alpha_i}
\right\}
\;\Longrightarrow\;
\hat{\alpha}_i \approx 
- \frac{\log\!\left(1 - \hat{\beta} \hat{t}_i \right)}
{\log\!\left(1 + \Delta^2 / \hat{\theta_i^2}\right)},
\end{equation}
where $\hat{\beta}$ and $\hat{\theta}_i$'s are obtained by maximum likelihood estimation, and $\hat{t}_i$'s are obtained directly from data using (\ref{eq:ti}). When $\hat{t}_i = 0$, the relationship in \eqref{eq:analytically solve alpha} implies that $\hat{\alpha}_i = 0$. In other words, when the input does not contribute to the variation of the output, it also does not have an impact on the correlation function in the GP surrogate. Such inactive inputs are automatically removed from the MIM kernel, allowing the GP model to be fitted to a reduced active subspace, which makes the maximum likelihood estimation easier.


Although the combination of the MOFAT design and the MIM kernel has been shown to outperform other existing methods, the improvement primarily comes from the MOFAT design's ability to estimate $\{\hat{t}_i\}_{i=1}^p$. In surrogate modeling, however, space-fillingness remains a fundamental requirement, as it ensures robust prediction over the input space. As discussed in the previous section, the MOFAT design does not guarantee optimal space-fillingness within the class of OFAT designs. Therefore, in the next section, we propose a new class of \textit{Space-filling One-Factor-at-a-Time} (SOFT) designs that optimize space-fillingness  and preserve the screening capability simultaneously.

\section{SOFT Designs}
\label{sec:SOFT Designs}
In general, a collection of $l$ OFAT designs can be defined as 
\begin{equation}
    \label{eq:general def for OFAT}
    \begin{aligned}
    \bm{D} =
    \begin{pmatrix} 
        \bm{O}^{1} \\
        \bm{O}^{2} \\
        \vdots \\
        \bm{O}^{l}
    \end{pmatrix},
    \; \text{where }
    \bm{O}^j
    =
    \begin{pmatrix}
        \bm{A}_j^{(0)} \\
        \bm{A}_j^{(1)} \\
        \vdots \\
        \bm{A}_j^{(p)}
    \end{pmatrix}
    \end{aligned}
    \xRightarrow{\text{rearrange the rows of } \bm{D}}
    \bm{D} =
    \begin{pmatrix} 
        \bm{A}^{(0)} \\
        \bm{A}^{(1)} \\
        \vdots \\
        \bm{A}^{(p)}
    \end{pmatrix}.
\end{equation}
Here, $\bm{A}^{(0)}$ is an $l\times p$ matrix representing the $l$ base runs. Without loss of generality, the columns of $\bm{D}$ can be arranged such that, for $i = 1, \dots, p$, $\bm{A}^{(i)}$ is obtained by changing the $i$th column of a previously generated matrix $\bm{A}^{(o_i)}$, where $o_i \in \{0, \dots, i-1\}$ indicates the index of the matrix from which $\bm{A}^{(i)}$ originates. The value of $o_i$ is determined by the type of OFAT designs. Under the standard structure, $o_i = 0$ for all $i$, whereas under the strict structure, $o_i = i-1$, if the factors are changed sequentially from 1 to $p$. Consequently, the space-fillingness of $\bm{D}$ is governed by three key elements:  
\begin{enumerate}
    \item[(i)] factor levels,
    \item[(ii)] OFAT structure, and
    \item[(iii)] the base runs $\bm{A}^{(0)}$.
\end{enumerate}
Our goal is to optimally choose these three elements so that $\bm{D}$ becomes as space-filling as possible. To guide this choice, we first introduce a suitable space-filling criterion.

\begin{figure}[ht!]
    \centering
    \includegraphics[scale=0.06]{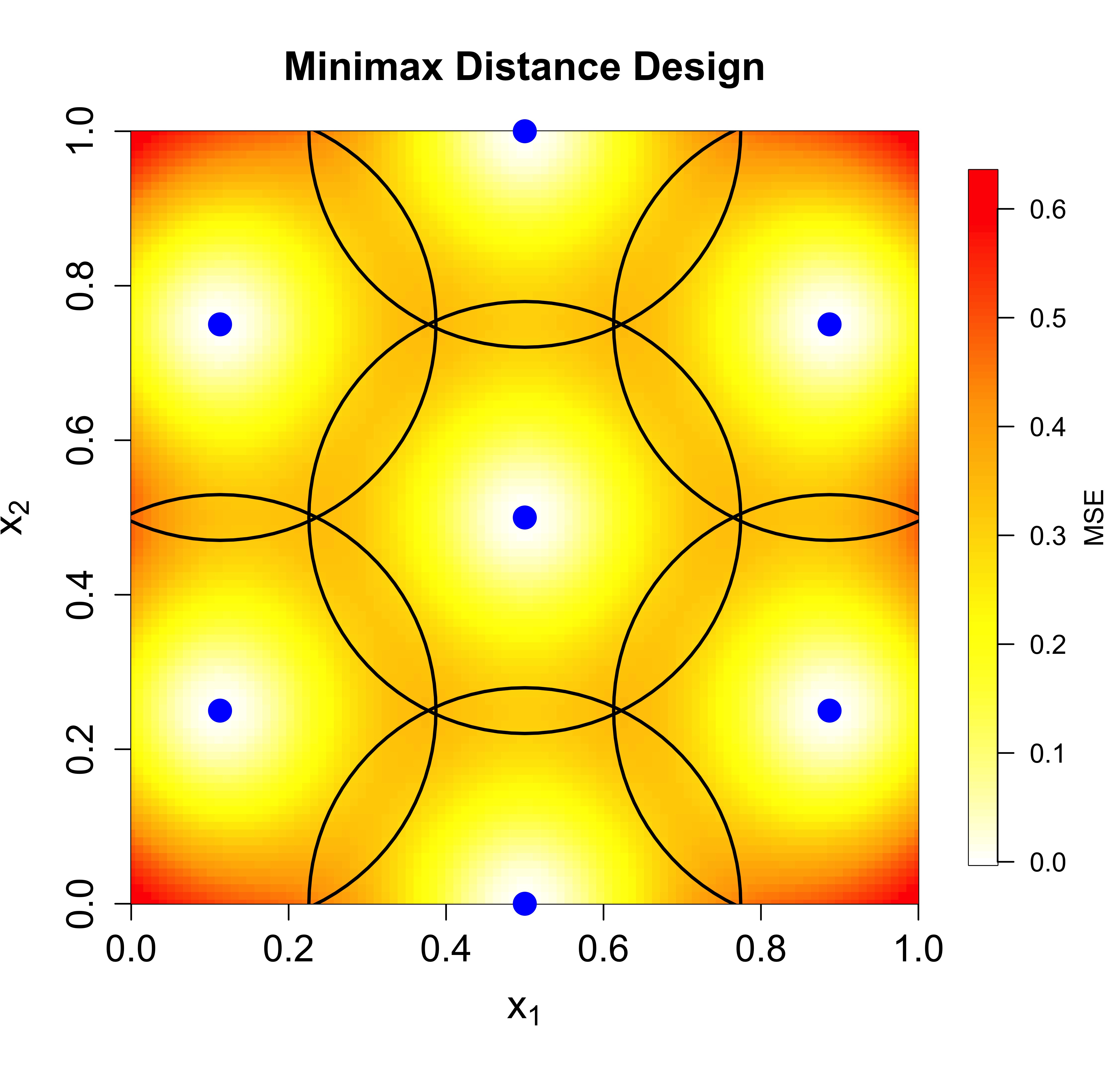}
    \caption{Minimax distance design for $n = 7$ points (solid blue circles). The MSE for a GP is calculated using an MIM kernel with $\alpha=1$ and $\theta=1/3$.}
    \label{fig:Minimax Distance Design}
\end{figure}

Minimax design \citep{johnson1990minimax} is one of the basic space-filling designs. It tries to minimize the maximum distance from all the points $\mathbf{x} \in \mathcal{X}$ to their closest neighbor in $\bm{D}$, which is obtained by minimizing the fill distance:
\begin{equation}
    \label{eq:minimax}
    \max_{\mathbf{x} \in \mathcal{X}} \min_{\mathbf{x}_i \in \bm{D}} \norm{\mathbf{x} - \mathbf{x}_i}_2.
\end{equation}
\citet{johnson1990minimax} has shown that the criterion minimizes the maximum mean squared error (MMSE, \citet{sacks1989design}) under a GP model with the near-independence assumption (i.e., any two distinct points are nearly uncorrelated). However, when the assumption does not hold, the points near the vertices (corners) of the input space tend to have a larger mean squared error (MSE) than those in the interior, as shown in Figure~\ref{fig:Minimax Distance Design}. This indicates that the criterion in \eqref{eq:minimax} does not account for the greater difficulty of extrapolation compared to interpolation.
Therefore, to better align the criterion with predictive performance, we consider directly minimizing the MMSE  under the GP surrogate introduced in Section~\ref{subsec:Gaussian Process Modeling} without the near-independence assumption:
\begin{equation}
    \label{eq:MMSE}
    MMSE(\bm{D}; \bm{\alpha}, \bm{\theta}) = \max_{\mathbf{x} \in \mathcal{X}} \{  1 - \mathbf{r}(\mathbf{x})' \mathbf{R}^{-1} \mathbf{r}(\mathbf{x}) \},
\end{equation}
where $\bm{\alpha} = \{\alpha_i\}_{i=1}^p$, $\bm{\theta} = \{\theta_i\}_{i=1}^p$, $\mathbf{r}(\mathbf{x}) = \{R(\mathbf{x}, \mathbf{x}_i)\}_{i=1}^n$ and $\mathbf{R} = \{R(\mathbf{x}_i, \mathbf{x}_j)\}_{i,j=1}^n$, with $n = l\times (p+1)$ denoting the run size. To avoid favoring any input dimension a priori, we assume $\alpha_i = \alpha$ and $\theta_i = \theta$ for all $i$. Under this assumption, the criterion simplifies to $MMSE(\bm{D}; \alpha, \theta)$. To move beyond the near-independence assumption while keeping the MMSE criterion simple, we fix $\alpha = 1$ and set $\theta = 1/m$, where all factors have $m$ levels. Minimizing MMSE is equivalent to maximizing the quadratic form
\begin{equation}
    \label{eq:mMSEc}
    Q(\bm{D}; \alpha, \theta) = \min_{\mathbf{x} \in \mathcal{X}} \{ \mathbf{r}(\mathbf{x})' \mathbf{R}^{-1} \mathbf{r}(\mathbf{x}) \},
\end{equation} 
which will be used as the space-filling criterion for optimizing the SOFT designs. Details on the numerical evaluation of \(Q\) are provided in the supplementary materials.

\subsection{Choice of Factor Levels}
\label{subsec:Choice of Factor Levels}
For the $i$th coordinate, $\bm{A}_{ji}^{(k)}$ takes the value of $\bm{A}_{ji}^{(o_i)}$ or $\bm{A}_{ji}^{(i)}$ for $k \neq \{o_i, i\}$. Thus, the levels of the $i$th coordinate are determined by the two matrices $\bm{A}^{(o_i)}$ and $\bm{A}^{(i)}$. Let $L_i = \{\bm{A}_{ji}^{(o_i)}, \bm{A}_{ji}^{(i)} \}_{j=1}^l$ denote the set of levels of $\bm{D}$ in the $i$th coordinate. Without prior information, we use the same level set for all factors, $L_1 = \cdots = L_p = L$. To achieve optimal one-dimensional projections, the levels in $L$ must be distinct, so that $L$ contains $2l$ elements. To obtain analytical expressions for the estimates $\{\hat{\alpha}_i\}_{i=1}^p$ under the MIM kernel as in (\ref{eq:analytically solve alpha}), the perturbation magnitude $\{|\bm{A}_{ji}^{(o_i)} - \bm{A}_{ji}^{(i)}|\}_{j=1}^l$ must be a constant $\Delta$. This leads to $2l$ equally-spaced levels: 
\begin{equation}
    \label{eq:levels}
    L = \{a, a + \delta, a + 2\delta, \dots, a + (2l-1)\delta\},
\end{equation}
where $a \in [0,0.5)$ and $\delta = (1-2a)/(2l-1)$. The levels are then paired to form the $l$ perturbations $\{(\bm{A}_{ji}^{(o_i)}, \bm{A}_{ji}^{(i)})\}_{j=1}^l$. To retain the screening capacity as in \eqref{eq:MOFAT criterion}, the total perturbation magnitude should be maximized,
\begin{align*}
    \{(\bm{A}_{ji}^{(o_i)}, \bm{A}_{ji}^{(i)})\}_{j=1}^l &= \argmax_{\{(\bm{A}_{ji}^{(o_i)}, \bm{A}_{ji}^{(i)})\}_{j=1}^l} \sum_{j=1}^l |\bm{A}_{ji}^{(o_i)} - \bm{A}_{ji}^{(i)}|.
\end{align*}
\begin{proposition}
    \label{prop: pairs of levels}
    Suppose $L$ are partitioned into $l$ disjoint pairs $(u_j, v_j)$, $j=1,\dots,l$ and the perturbation magnitudes satisfy 
$|u_j - v_j| = \Delta
\quad \text{for all } j$.
Then, $
\sum_{j=1}^l |u_j - v_j|$
is maximized when $\Delta = l\delta$, which is achieved by pairing
\[
(u_j, v_j) = \bigl(a + k\delta,\; a + (k+l)\delta\bigr),
\qquad k=0,1,\dots,l-1.
\]
\end{proposition}

All the proofs are given in the supplementary materials. From Proposition~\ref{prop: pairs of levels}, the optimal pairing matches each level with the one located $l\delta$ distance away. This leads to the following transformation for obtaining $\bm{A}_{ji}^{(i)}$,
\begin{equation}
    \label{eq:our transformation}
    \bm{A}_{ji}^{(i)} = T_l^*(\bm{A}_{ji}^{(o_i)}), \quad
    T_l^*(x) = x \pm l\delta,
\end{equation}
where the sign is chosen such that $T_l^*(x) \in [0,1]$. Interestingly, the transformation is an involution:
$
T_l^*(T_l^*(\bm{A}_{ji}^{(o_i)})) = \bm{A}_{ji}^{(o_i)}$.
A graphical illustration of the construction leading to the transformation in \eqref{eq:our transformation} is provided in the supplementary materials.
If $l$ is even and $L$ is chosen as $\{0.25/l, 0.75/l, \dots, 1-0.25/l\}$, then the transformation in \eqref{eq:our transformation} coincides with the MOFAT transformation in \eqref{eq:mofat transformation}. However, this choice of $L$ does not necessarily lead to a space-filling design $\bm{D}$.


\begin{proposition}
    \label{prop:one-dimensional mMSEc}
    Suppose a product correlation function is used in the GP surrogate. Let $L_i$ denote the set of $2l$ levels of $\bm{D}$ in the $i$th coordinate. Then maximizing $\prod_{i=1}^p q(L_i; \alpha_i, \theta_i)$ raises the lower and upper bounds of $Q(\bm{D}; \bm{\alpha}, \bm{\theta})$, where 
    \[
    q(L_i;\alpha_i,\theta_i)=\min_{x \in [0,1]} \{ \mathbf{r}_i(x)' \mathbf{R}_i^{-1} \mathbf{r}_i(x) \},
    \]
    with $\mathbf{r}_i(x) = \{ R(x, L_{ij}) \}_{j=1}^{2l}$ and $\mathbf{R}_i = \{R(L_{ij}, L_{ik})\}_{j,k=1}^{2l}$.
\end{proposition}

From Proposition~\ref{prop:one-dimensional mMSEc}, to maximize $Q(\bm{D}; \alpha, \theta)$, we should also maximize $q(L; \alpha, \theta)$. Because $L$ is fully determined by the first element $a$, the optimization reduces to a one-dimensional problem:
\[
a^* = \argmax_a q(L; \alpha, \theta).
\]
However, this optimization depends on the unknown hyperparameters $\alpha$ and $\theta$, which control the correlation between design points. When $\alpha$ and $\theta$ are selected such that the design points are nearly independent, the MMSE criterion reduces to the fill distance criterion and we obtain $a^* = 0.25/l$. In contrast, when the design points are highly correlated, the prediction error tends to increase near the boundary of the input space, and the optimal choice becomes $a^* = 0$. Therefore, instead of optimizing for $a^*$ directly, we choose a fixed value of $a$ that is robust across different possible values of $\alpha$ and $\theta$. Since $0.25/l$ and $0$ correspond to two extreme correlation settings, the midpoint setting $a = 0.125/l$ could be a robust choice, which is investigated next.

\begin{figure}[ht!]
    \centering
    \includegraphics[scale=0.175]{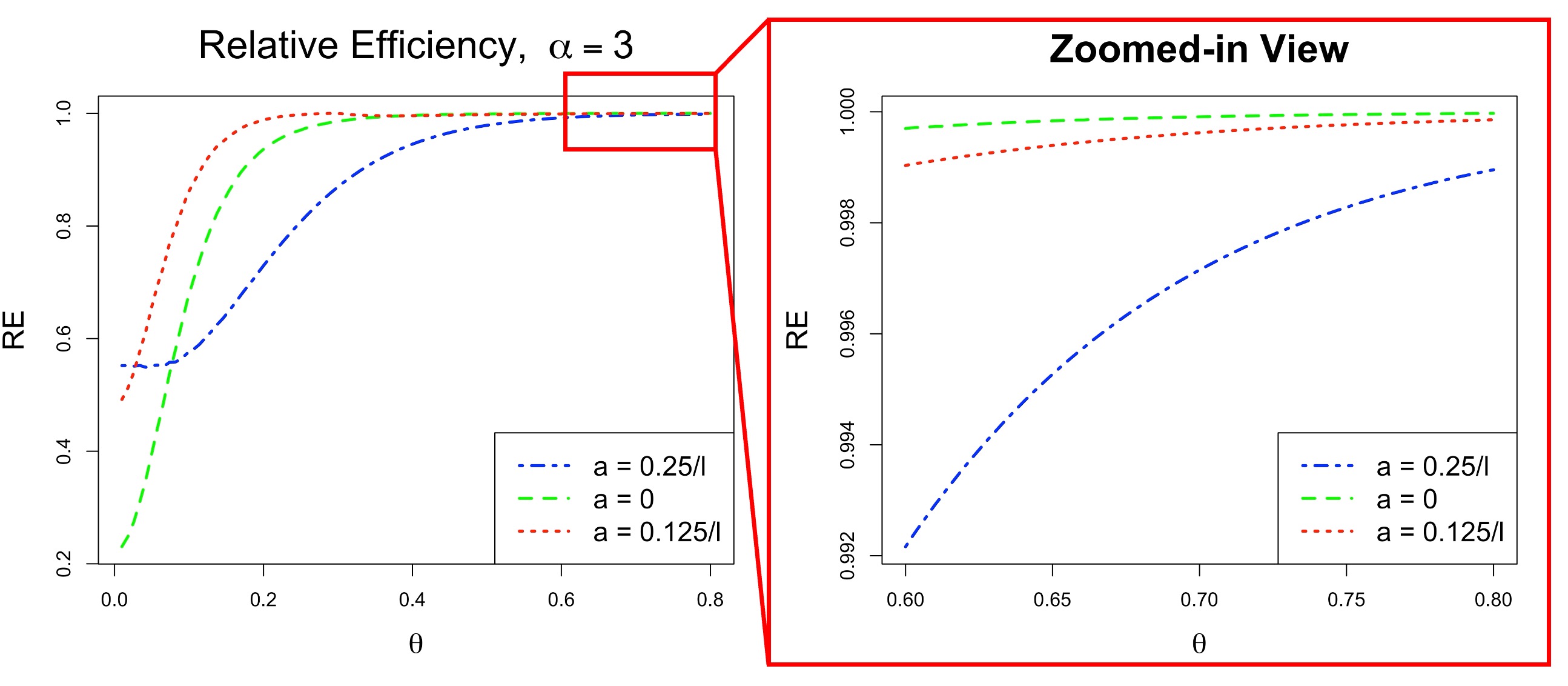}
    \caption{Relative efficiency of the three choices of $a$ when $l = 4$.}
    \label{fig:choice of a 3}
\end{figure}

Figure~\ref{fig:choice of a 3} illustrates the relative efficiency of the three choices of $a$ under different correlation settings. The relative efficiency (RE) is defined as \citep{sacks1989designs} 
\[
    RE(L; \alpha, \theta) = \frac{q(L; \alpha, \theta)}{\max_L q(L; \alpha, \theta)}.
\]
When $\alpha = 3$ and $\theta$ is small (low correlation), $a = 0.25/l$ achieves the highest RE, followed by $a = 0.125/l$, while $a = 0$ performs poorly. However, as $\theta$ increases, $a = 0.125/l$ quickly becomes the best option, followed by $a = 0$, while $a = 0.25/l$ performs the worst. When $\theta$ is large (strong correlation), the zoomed-in panel shows that $a = 0$ achieves the highest RE, followed by $a = 0.125/l$, and then $a = 0.25/l$. Although the optimal choice of $a$ varies with the correlation, $a = 0.125/l$ consistently maintains relatively high RE throughout the range of $\theta$, making it a robust choice.

Regardless of the OFAT structure, there is an imbalance in the frequency between any base level $\bm{A}_{ji}^{(o_i)}$ and its paired level $\bm{A}_{ji}^{(i)}$. To prevent the design points from concentrating toward one side of the input space and to make the design more balanced, both the base and paired levels should be arranged symmetrically within $[0,1]$, which is only possible when $l$ is even. Therefore, we restrict our attention to even  $l$ throughout this work.


\subsection{OFAT Structures}
\begin{figure}[ht!]
    \centering
    \includegraphics[scale=0.083]{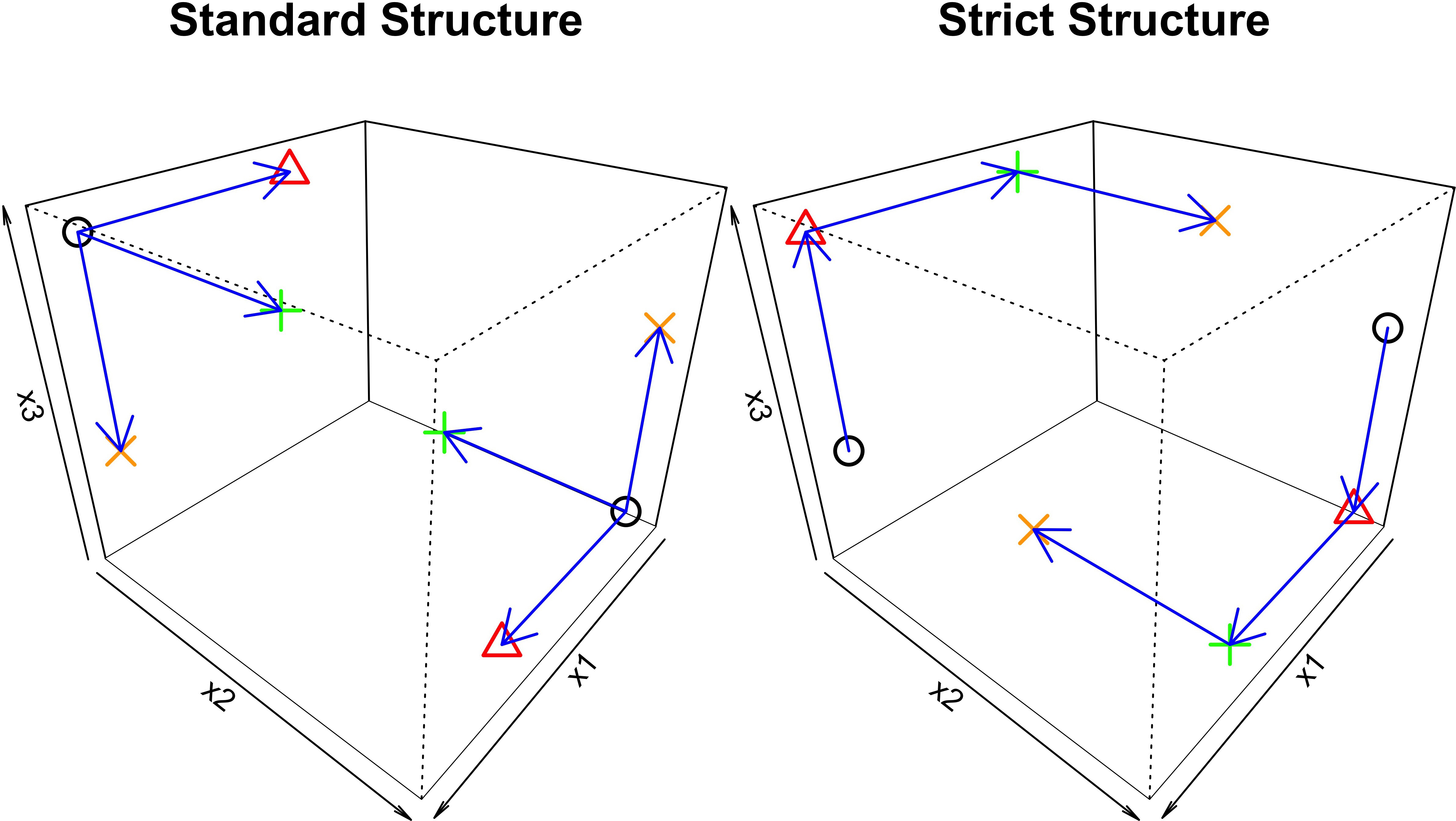}
    \caption{OFAT designs under the standard and strict structures with $p = 3$ and $l = 2$. The strict structure uses the order $\bm s=(3,1,2)$.}
    \label{fig:three_panel}
\end{figure}

Recall that $\bm{A}^{(i)}$ is obtained by changing the $i$th column of $\bm{A}^{(o_i)}$, where $o_i$ is determined by the OFAT structure. See Figure~\ref{fig:three_panel} for an illustration. Under the standard structure, $o_i = 0$ for all $i$. Therefore, given the transformation in \eqref{eq:our transformation}, the resulting design is uniquely determined by $\bm{A}^{(0)}$. We express the design under the standard structure as
\begin{equation}
    \label{eq:standard}
    \bm{D} = f_{stand}(\bm{A}^{(0)}) \coloneqq (\bm{A}^{(0)^T}, \bm{A}^{(1)^T}, \dots, \bm{A}^{(p)^T})^T.
\end{equation}

In contrast, the design under the strict structure depends not only on $\bm{A}^{(0)}$, but also on the order in which the factors are changed. For example, if the factors are changed sequentially from 1 to $p$, then $o_i = i - 1$. Let $\bm{s} = (s_1, \dots, s_p)$ be a permutation of $\{1, \dots, p\}$, where $s_k$ denotes the index of the factor changed at step $k$. The first change $s_1$ is based on $\bm{A}^{(0)}$ and $o_{s_{k}} = s_{k-1}$ for $k \geq 2$. Therefore, the design under the strict structure can be expressed as
\begin{equation}
    \label{eq:strict}
    \bm{D} = f_{strict}(\bm{A}^{(0)}, \bm{s}) \coloneqq (\bm{A}^{(0)^T}, \bm{A}^{(s_1)^T}, \dots, \bm{A}^{(s_p)^T})^T,
\end{equation}
where $\bm{A}^{(s_k)}$ is obtained by changing the $s_k$th column of $\bm{A}^{(s_{k-1})}$ using the transformation in \eqref{eq:our transformation}. 
To obtain the optimal space-filling OFAT design under the strict structure, one would need to examine $p!$ possible orders corresponding to the permutations of $\{1, \dots, p\}$. Since this is computationally infeasible for large $p$, we instead propose a greedy algorithm to construct the design.

Let $S_i$ denote a sequence of matrices that records the construction history of the design after changing the $i$ factors, for $i = 0, \dots, p$. For $i \geq 1$, any two consecutive elements in $S_i$ differ exactly in one column. Initially, $S_0 = (\bm{A}^{(0)})$. At step $(i+1)$, the $(i+1)$th factor is changed from either the first or the last element of $S_i$. This produces two candidate sequences by concatenation: 
\[
S_{i+1}^1 = (\bm{A}^{(i+1)}, S_{i}), \quad S_{i+1}^2 = (S_{i}, \bm{A}^{(i+1)}).
\]
For each candidate sequence, a corresponding design is formed by stacking its elements row-wise. The two designs are then evaluated using $Q$ and the corresponding sequence with the larger value is selected and indicated by $S_{i+1}$. After $p$ steps, the design $\bm{D}$ is constructed by stacking the elements of $S_p$ row by row. The procedure is summarized in Algorithm~\ref{alg:Greedy Algorithm for Strict Structure}. 
\begin{algorithm}
\caption{Greedy Algorithm for Strict Structure}
\label{alg:Greedy Algorithm for Strict Structure}
\renewcommand{\baselinestretch}{1.3}\selectfont
\begin{algorithmic}[1]
    \Require LHD $\bm{A}^{(0)}$.
    \State Set $\alpha = 1$, $\theta = 1/(2l)$.
    \State Initialize $S_0 = (\bm{A}^{(0)})$.
    \State Set $S_1 = (\bm{A}^{(0)}, \bm{A}^{(1)})$.
    \For{$i = 1, \dots, (p-1)$}
        \State Construct $S_{i+1}^1 = (\bm{A}^{(i+1)}, S_{i})$ and $S_{i+1}^2 = (S_{i}, \bm{A}^{(i+1)})$.
        \State Form $\bm{D}_{i+1}^1$ and $\bm{D}_{i+1}^2$ by stacking $S_{i+1}^1$ and $S_{i+1}^2$ row-wise.
        \If{$Q(\bm{D}_{i+1}^1; \alpha, \theta) > Q(\bm{D}_{i+1}^2; \alpha, \theta)$}
            \State Set $S_{i+1} = S_{i+1}^1$.
        \Else
            \State Set $S_{i+1} = S_{i+1}^2$.
        \EndIf
    \EndFor
    \State Form $\bm{D}$ by stacking $S_p$ row-wise.
    \State \Return $\bm{D}$.
\end{algorithmic}
\end{algorithm}

\subsection{Optimization of Base Runs}
Regardless of the OFAT structure, the resulting design depends on the base runs $\bm{A}^{(0)}$. To ensure a balanced design, the set of levels used to construct $\bm{A}^{(0)}$ is specified as
\begin{equation}
   \label{eq:set of base levels}
   L^0 = \{L'_{2k-1}\}_{k=1}^{l/2} \cup \{L'_{l+2k}\}_{k=1}^{l/2},
\end{equation}
where $L'$ is a sequence of elements of $L$ arranged in increasing order. For example, if $l = 4$, then $L^0 = \{a, a+2\delta, a+5\delta, a+7\delta\}$.
Now, we can obtain the optimal $\bm{A}^{(0)}$ for the standard version by
\[\argmax_{\bm A^{(0)}} Q(f_{stand}(\bm A^{(0)}); \alpha,\theta)\]
and the strict version by
\[\argmax_{\bm A^{(0)}} Q(f_{strict}(\bm A^{(0)},\bm s^*(\bm A^{(0)}));\alpha,\theta),\]
where $\bm s^*(\bm A^{(0)})$ is obtained using Algorithm 1.
We propose using a deterministic local search algorithm for optimization \citep{wang2026sfdesign}, where a pair of design points in each coordinate is swapped. The swap is accepted only if it leads to an improvement of $Q$. Further details of the deterministic local search algorithm are provided in the supplementary materials.

However, if the number of coordinates $p$ is large, the algorithm will be computationally prohibitive. To reduce the computational burden, we adopt the following simplification proposed in \citet{xiao2023maximum}. Let $\bm{P}_{\bm{A}}$ be the $l \times l!$ matrix whose columns are all possible permutations of $L^0$.
When $p \geq l!$, we construct $\bm{A}^{(0)}$ by repeating $\bm{P}_{\bm{A}}$ for $\lfloor p/(l!) \rfloor$ times, so that each permutation appears approximately the same number of times, resulting in a more balanced LHD. Since $\bm{P}_{\bm{A}}$ already contains every permutation, applying the local search algorithm to these replicated blocks does not improve the objective due to symmetry. Therefore, the deterministic local search needs only to be applied to the remaining $(p \mod l!)$ columns of $\bm{A}^{(0)}$. The algorithm for generating a space-filling OFAT (SOFT) design is provided in Algorithm~\ref{alg:SOFT Design}. An illustration of the SOFT design under the standard structure with $p = 2$ and $l = 6$ is provided in Figure~\ref{fig:previous OFATs}(d). We can clearly see that this design provides the best space-fillingness among the designs shown in Figure~\ref{fig:previous OFATs}.


\begin{algorithm}
\caption{SOFT Design}
\label{alg:SOFT Design}
\renewcommand{\baselinestretch}{1.3}\selectfont
\begin{algorithmic}[1]
    \Require Number of factors $p$, number of OFAT sizes $l$, OFAT structure.
    \State Set $L = \{\frac{0.125}{l}, \frac{0.125}{l} + \delta, \dots, \frac{0.125}{l} + (2l-1)\delta\}$, where $\delta = \frac{l-0.25}{l(2l-1)}$.
    \State Set $L^0 = \{L'_{2k-1}\}_{k=1}^{l/2} \cup \{L'_{l+2k}\}_{k=1}^{l/2}$, where $L'$ is a sequence of elements of $L$ arranged in increasing order.
    \If{$p \geq l!$}
        \State Initialize $\bm{P}_{\bm{A}}$ with levels in $L^0$.
        \State Initialize $\bm{A}^{(0)}$ as $\lfloor p/(l!) \rfloor$ replicates of $\bm{P}_{\bm{A}}$ plus its first $(p \mod l!)$ columns.
        \State Set $\bm{A}^{(0)'}$ to be the last $(p \mod l!)$ columns of $\bm{A}^{(0)}$.
    \Else 
        \State Initialize $\bm{A}^{(0)}$ as an MmLHD with levels in $L^0$.
        \State Set $\bm{A}^{(0)'} = \bm{A}^{(0)}$.
    \EndIf
    \State Apply the deterministic local search algorithm to $\bm{A}^{(0)'}$.
    \State \Return $\bm{D}$.
\end{algorithmic}
\end{algorithm}


\section{Numerical Studies}
\label{sec:Numerical Studies}
In this section, we evaluate the performance of SOFT designs under the standard (SOFT-Stand) and strict (SOFT-Strict) structures against MmLHD, MaxPro and MOFAT designs. The MmLHD and MaxPro designs are generated using the R package \texttt{SLHD} \citep{Ba2015SLHD} and \texttt{MaxPro} \citep{Ba2018maxpro}, respectively. The MOFAT designs are implemented using the R package \texttt{MOFAT} \citep{Xiao2022mofat}, which uses an MmLHD for its base runs $\bm A^{(0)}$. To ensure a fair comparison, we set the run size of the MmLHD and MaxPro designs to $n = l(p+1)$, consistent with the OFAT-type designs.

\subsection{Space-Filling Properties}
\label{subsec:Space-Filling Properties}

To evaluate space-filling properties, we first consider the proposed criterion $Q(\bm{D}; \alpha, \theta)$ with $\alpha = 1$ and $\theta = 1/2l$ in the MIM kernel, where larger values of $Q$ indicate better space-filling performance. The evaluation of $Q$ is computed over 10,000 Sobol' sequences generated using the R package \texttt{spacefillr} \citep{Morgan-Wall2025spacefillr} and the corresponding vertex points in the input space; details on the construction of the evaluation set are provided in Section~S1 of the supplementary materials.
In addition, we consider the criterion in \citet[p.~54]{joseph2026experimental}, defined as
\begin{equation}
    \label{eq:entropy-based criterion}
    \min_{\bm{D}} \Phi = \min_{\bm{D}} \sum_{i=1}^n \sum_{j \neq i} \frac{1}{ 1 + \norm{\mathbf{x}_i - \mathbf{x}_j}^2 / \gamma^2},
\end{equation}
which is closely related to the maximum entropy \citep{shewry1987maximum} and the total reciprocal distance \citep{morris1995exploratory} criteria. The hyperparameter is set to $\gamma=\sqrt{p}/(2l)$, which preserves the coordinatewise lengthscale $1/(2l)$ used in $Q$. The designs are compared across different numbers of factors $p = 2, \dots, 20$ with $l = 6$. For each $p$, the simulation is repeated 30 times.

\begin{figure}[ht!]
    \centering
    \includegraphics[scale=0.73]{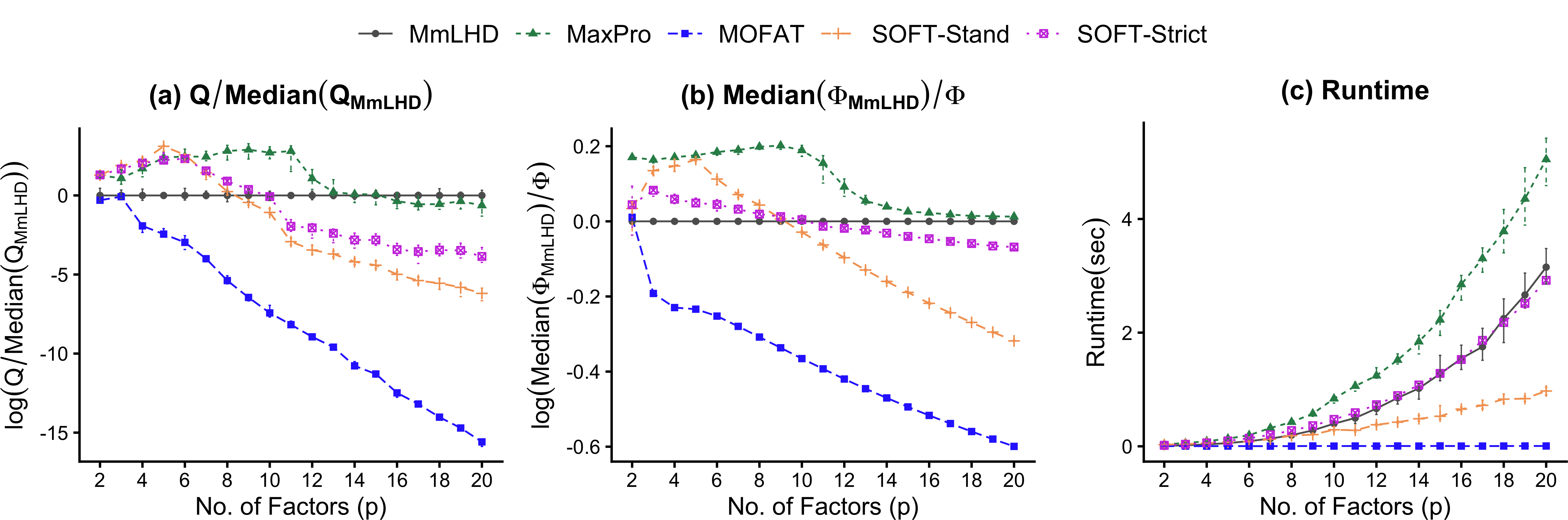}
    \caption{Comparison of designs in $Q$, $\Phi$, and runtime across different numbers of factors. For each $p$, $\operatorname{Median}(Q_{\mathrm{MmLHD}})$ and $\operatorname{Median}(\Phi_{\mathrm{MmLHD}})$ denote the median $Q$ and $\Phi$ values of MmLHD over 30 independent replications.}
    \label{fig:combined}
\end{figure}

Figure~\ref{fig:combined}(a) shows the log ratio of $Q$ to the median MmLHD value at each $p$. As expected, space-filling designs such as MmLHD and MaxPro consistently produce large $Q$ across all values of $p$. SOFT designs achieve performance comparable to these space-filling designs for $p \leq 10$ and still maintain stable $Q$ values as $p$ increases, demonstrating their ability to preserve good space-filling properties in high-dimensional settings. In contrast, MOFAT designs consistently produce the smallest $Q$ and their performance deteriorates rapidly as $p$ increases, indicating poor space-filling performance.

Figure~\ref{fig:combined}(b) shows the log ratio of the median MmLHD $\Phi$ value to each design's $\Phi$. Since smaller $\Phi$ is preferred, higher values of $\log(\text{Median} (\Phi_{\text{MmLHD}}) / \Phi)$ indicate better space-filling performance. Overall, MaxPro performs best across all values of $p$. For SOFT designs, both variants consistently outperform MOFAT and can even surpass MmLHD in lower-dimensional settings, but their performance declines as $p$ increases. Taken together with the results in Figure~\ref{fig:combined}(a), these findings demonstrate that SOFT designs achieve strong space-filling performance across different criteria. They remain competitive with classical space-filling designs while substantially outperforming MOFAT, especially in high-dimensional settings. Between the two SOFT variants, SOFT-Strict generally performs better than SOFT-Stand.

In addition to space-filling performance, we examine the computational cost of constructing these designs. All experiments are conducted on a Mac mini equipped with an M4 chip. As shown in Figure~\ref{fig:combined}(c), MaxPro incurs the highest computational cost. MmLHD and SOFT-Strict have similar runtimes, whereas SOFT-Stand is considerably more efficient. This difference reflects the computational complexity associated with generating the two different OFAT structures. MOFAT designs require negligible computational cost, as they are constructed without optimization.

\subsection{Factor Screening}
\label{subsec:Factor Screening}
In this section, we evaluate the factor-screening performance of each design using the total Sobol' indices $\{t_i\}_{i=1}^p$. For the OFAT-type designs, including MOFAT and SOFT, these indices are estimated directly from the data using Equation~\eqref{eq:ti}. For the space-filling designs, MmLHD and MaxPro, we first
fit a Gaussian process model with the Gaussian kernel:
\begin{equation}
\label{eq:gaussian_kernel}
R(\mathbf{x},\mathbf{x}')
=
\exp\left(
-\sum_{i=1}^{p}
\frac{(x_i-x_i')^2}{\theta_i^2}
\right),
\end{equation}
using the R package \texttt{rkriging}
\citep{Huang2025rkriging}. The total Sobol' indices are then estimated from the fitted model using the R package \texttt{sensitivity}
\citep{Iooss2026sensitivity}.

For this study, we consider the g-function, which originally has eight inputs. See the supplementary materials for details. To create a more challenging screening problem, we augment the function with 12 weakly influential inputs, resulting in a total dimension of $p=20$. The added inputs contribute through both main effects and
interactions. Specifically, let $g(\mathbf{x})$ denote the original benchmark function with
$\mathbf{x}=(x_1,\ldots,x_{p_0})$, and let
$\mathbf{z}=(z_1,\ldots,z_{p_1})$ denote the additional inputs. The augmented function is defined as
\begin{equation}
g^{\mathrm{aug}}(\mathbf{x},\mathbf{z})
=
g(\mathbf{x})
+
\sum_{i=1}^{p_1}\beta_i z_i
+
\sum_{i=1}^{p_1}\sum_{j=1}^{p_0}
\gamma_{ij}z_i x_j,
\label{eq:augmented_function}
\end{equation}
where the coefficients are independently generated from 
$
\beta_i,\gamma_{ij}
\overset{iid}{\sim}
N\left(0,\sigma_g^2/100\right)$,
and $\sigma_g^2$ is the variance of the original function output. The coefficients are generated once using a random seed and then held fixed throughout the simulation, so that $g^{\mathrm{aug}}$ remains deterministic. For each design, the run size is set to $8(p+1)$, and the
simulation is repeated 30 times. In each repetition, the assignment of factors to the columns of a design is randomized. To facilitate comparison, we normalized the estimated total Sobol' indices as
\[
\tilde{t}_i = \frac{\hat{t}_i}{\sum_{i=1}^p \hat{t}_i}.
\]

\begin{figure}[h!]
    \centering
    \includegraphics[scale=0.115]{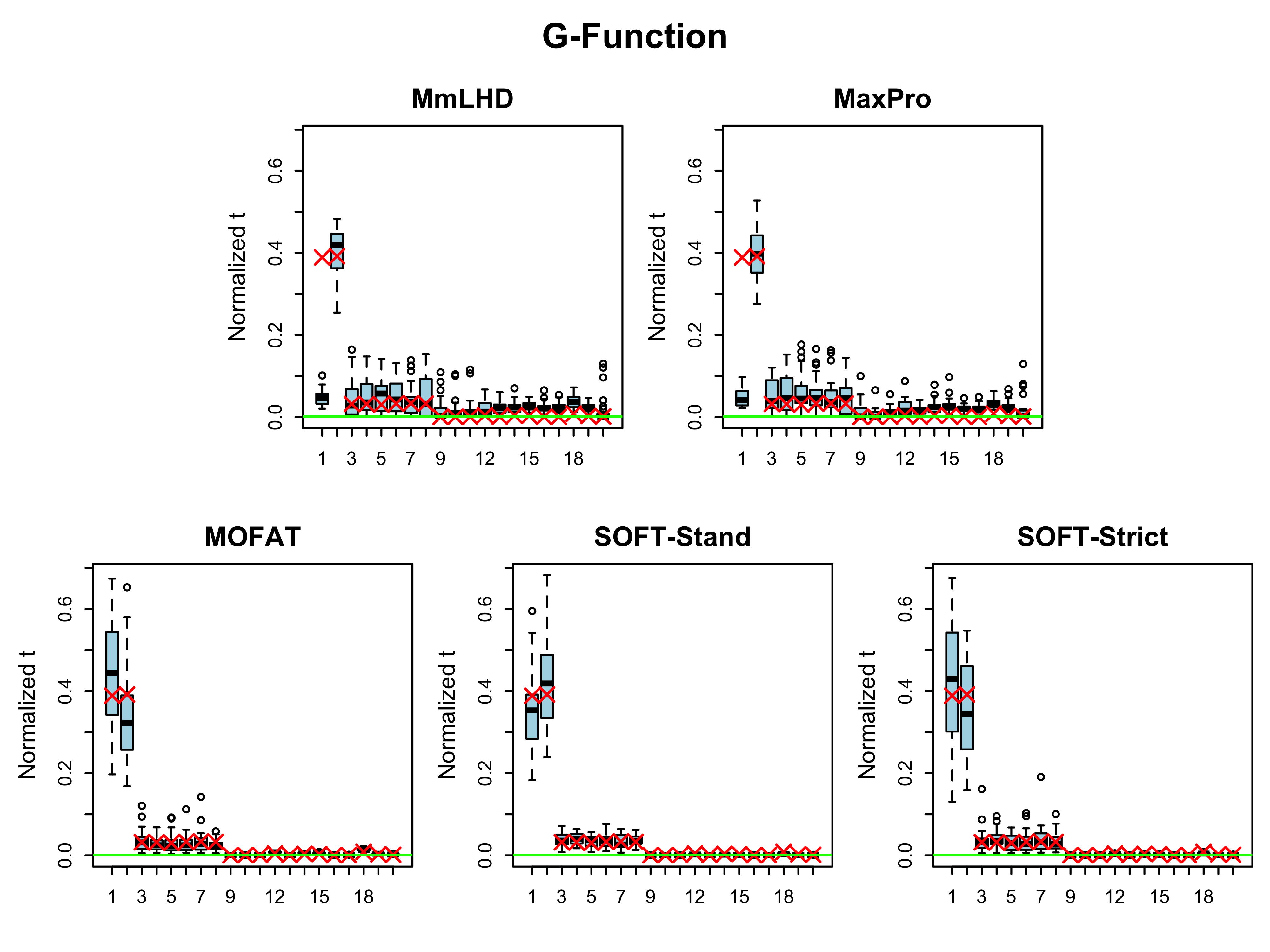}
    \caption{Boxplots of $\tilde{t}_i$ for the G-function. Red crosses indicate the reference normalized total Sobol' indices. The green horizontal line marks the screening threshold of $10^{-3}$, below which a factor is screened out.}
    \label{fig:t_gfun}
\end{figure}

The results are shown in Figure~\ref{fig:t_gfun}. The red crosses represent
the reference normalized total Sobol' indices computed using a 10,000-point Sobol' design defined in \eqref{eq:sobol' design}. The green horizontal line denotes the screening threshold of $10^{-3}$. Factors whose $\tilde{t}_i$ fall below this threshold are classified as negligible. For MmLHD and MaxPro, the estimated indices of many weakly influential factors remain above the threshold, indicating that these factors cannot be reliably screened out. In addition, both designs substantially underestimate the total Sobol' index of the first factor. Consequently, the estimated indices from MmLHD and MaxPro can give a misleading representation of the relative factor importance, making factor ranking unreliable. In contrast, MOFAT and both SOFT variants assign the indices of negligible factors below the screening threshold in most repetitions, illustrating substantially better screening performance. Moreover, the median estimated indices from SOFT are close to their reference values, showing that SOFT can also be used for factor ranking. Overall, the screening performance of SOFT is comparable to that of MOFAT. This demonstrates the improved space-fillingness of SOFT does not come at the expense of its factor-screening capability. Additional results for other benchmark functions are provided in the supplementary materials.

\subsection{Surrogate Modeling}
\label{subsec:Surrogate Modeling}
In this section, we assess the effectiveness of each design for fitting a surrogate model. We use the Gaussian process (GP) model with the MIM kernel introduced in Section~\ref{subsec:Gaussian Process Modeling}. Instead of estimating the full set of hyperparameters $\{\alpha_i\}_{i=1}^p$ and $\{\theta_i\}_{i=1}^p$, we first perform factor screening and subsequently fit the GP using only the factors identified as active.

For each design, the total Sobol' indices are estimated using the same procedure described in Section~\ref{subsec:Factor Screening}. Specifically, for MOFAT and SOFT, the indices are estimated directly from the OFAT structure, whereas for MmLHD and MaxPro, they are estimated from a preliminary GP model fitted with the Gaussian kernel.  Factors with normalized total Sobol' index $\tilde{t}_i < 10^{-3}$ are excluded from the subsequent surrogate modeling. Let $p'$ denote the number of factors retained after screening. The resulting GP therefore involves $\beta$ and $\{\theta_i\}_{i=1}^{p'}$ for MOFAT and SOFT, and $\{\alpha_i\}_{i=1}^{p'}$ and $\{\theta_i\}_{i=1}^{p'}$ for MmLHD and MaxPro. All remaining hyperparameters are estimated by maximum likelihood using a derivative-free local optimization algorithm implemented in the \texttt{NLopt} package \citep{NLopt}. The parameters $\alpha_i$ and $\theta_i$ are initialized at 1 and are bounded in $[0.1, 10]$ and $[0.01, 100]$, respectively. For the OFAT-type designs, the parameter $\beta$ is initialized at $\max_k (0.5/\tilde{t}_k)$ and is bounded in $[0.05, \max_k (1/\tilde{t}_k)-10^{-4}]$.



\begin{figure}[ht!]
    \centering
    \includegraphics[scale=0.06] {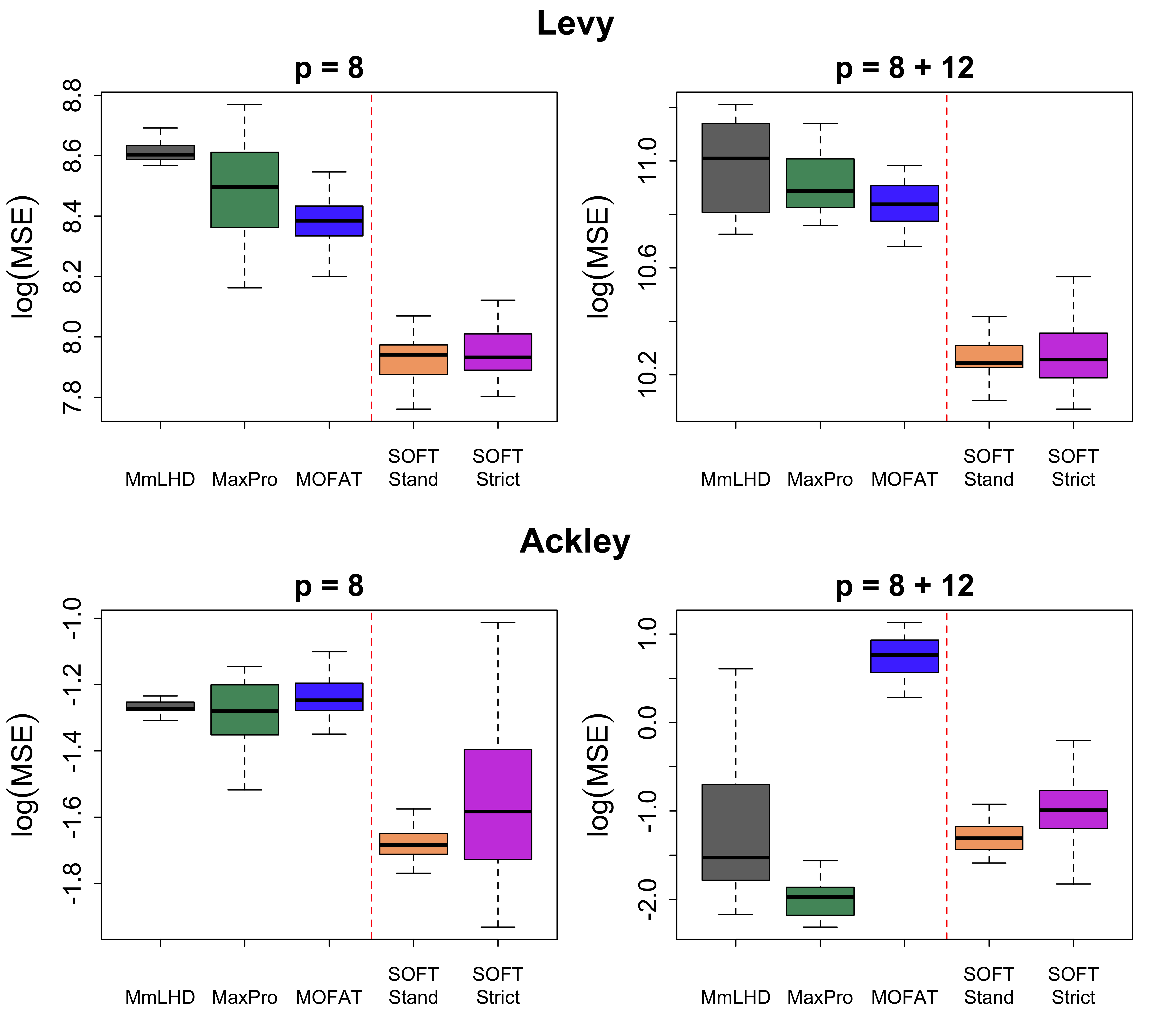}
    \caption{Out-of-sample log MSE for the Levy and Ackley functions under varying input dimensions.}
    \label{fig:all_mse_panels}
\end{figure}

We consider two benchmark functions, the Levy function and the Ackley function, both of which originally have eight inputs. To evaluate the performance in high-dimensional settings, we augment the original function with 12 inert inputs using the same construction as in (\ref{eq:augmented_function}), resulting in a 20-dimensional new problem. For each setting, the run size is set to $8(p+1)$, which is close to the recommended $10p$ run size in the literature \citep{loeppky2009choosing}. The simulations are repeated 30 times, where in each repetition, the assignment of factors to the columns of design is randomized. The performance of surrogate modeling is evaluated using the out-of-sample mean squared error (MSE), computed over the same Sobol'-based evaluation set described in Section~\ref{subsec:Space-Filling Properties}. The results for the numerical test functions are shown in Figure~\ref{fig:all_mse_panels}.

For the Levy function, the SOFT designs clearly achieve better performance than the space-filling designs (MmLHD and MaxPro) and MOFAT design for both 8 and 20 dimensional settings. On the other hand, the Ackley function exhibits a different pattern. When $p = 8$, SOFT-Stand achieves the lowest MSE, while SOFT-Strict also performs well but shows greater variability. However, after adding 12 inert factors, the performance of MmLHD and MaxPro improves and surpasses that of the SOFT designs, whereas MOFAT deteriorates substantially. This contrasting behavior can be attributed to the complexity of the Ackley function, whose structure has a broad, relatively flat outer region with local oscillations and a sharp central basin. Accurately approximating such a complex response surface requires substantial sampling effort. When $p$ increases from 8 to 20, the total run size increases from $8(8+1)=72$ to $8(20+1)=168$. For MmLHD and MaxPro, these additional runs provide more distinct points over the 8-dimensional active subspace. For OFAT-type designs, however, runs associated with perturbing the 12 inert factors do not introduce new points in this subspace, so the number of distinct points remains 72. This limitation arises from the OFAT structure itself rather than the space-filling construction of SOFT. Although factor screening mitigates the adverse effect of inert factors, the OFAT structure limits how the design points can be used. For the Ackley function, the benefit of unrestricted space-filling designs appears to outweigh the advantage of screening. This observation motivates the space-filling augmentation strategy considered in the next section, which combines SOFT with unrestricted space-filling points.



\section{Augmented SOFT Designs}
\label{sec:Augmented SOFT Designs}
Instead of allocating the entire experimental budget to a SOFT design, we allocate the budget between a SOFT component and a space-filling augmentation. The resulting design can be written as
\[
\bm{D}
=
\begin{pmatrix}
\bm{D}_{\mathrm{soft}}\\
\bm{D}_{\mathrm{aug}}
\end{pmatrix},
\]
where $\bm{D}_{\mathrm{soft}}$ is a SOFT design containing $n_{\mathrm{soft}} = l(p+1)$ runs, and $\bm{D}_{\mathrm{aug}}$ contains $n_{\mathrm{aug}}$ space-filling points. We refer to
$\bm{D}$ as an \textit{augmented SOFT design}, abbreviated as \textit{A-SOFT}. Consistent with the construction of SOFT designs in Section~\ref{sec:SOFT Designs}, the space-fillingness of $\bm{D}$ is optimized using the $Q$ criterion. To simplify the optimization, we adopt a two-stage approach: $\bm{D}_{\mathrm{soft}}$ is first constructed and held fixed, and the augmentation points in $\bm{D}_{\mathrm{aug}}$ are then selected sequentially. 

In Section~\ref{subsec:Choice of Factor Levels}, the first element of the level set $L$ in a SOFT design is chosen as $a = 0.25/(2l)$, where $2l$ is the number of levels in each factor. Since $\bm{D}$ can contain up to $m = 2l + n_{\mathrm{aug}}$ levels, this choice is extended to $a_{\mathrm{aug}} = 0.25/m$. The SOFT component $\bm{D}_{\mathrm{soft}}$ is then constructed as in Section~\ref{sec:SOFT Designs}, with $a_{\mathrm{aug}}$ replacing $a$. To preserve the boundary in levels after augmentation, the augmentation points are restricted to $\mathcal{X}_{\mathrm{aug}} = [a_{\mathrm{aug}}, 1 - a_{\mathrm{aug}}]^p$. With these choices, the augmentation points are selected sequentially. \\
Let $\bm{D}^{(0)} = \bm{D}_{\mathrm{soft}}$. At step $j = 1, \dots, n_{\mathrm{aug}}$, the augmentation point is selected as
\begin{equation}
    \label{eq:augmentation}
    \mathbf{x}_j^* = \arg\max_{\mathbf{x} \in \mathcal{X}_{\mathrm{aug}}} Q\left(
    \begin{pmatrix}
        \bm{D}^{(j-1)}\\
        \mathbf{x}
    \end{pmatrix}; \alpha = 1, \theta = \frac{1}{m}
    \right),
\end{equation}
and $\bm{D}^{(j)}$ is obtained by appending $\mathbf{x}_j^*$ to $\bm{D}^{(j-1)}$. 

\begin{figure}[ht!]
    \centering
    \includegraphics[scale=0.06]{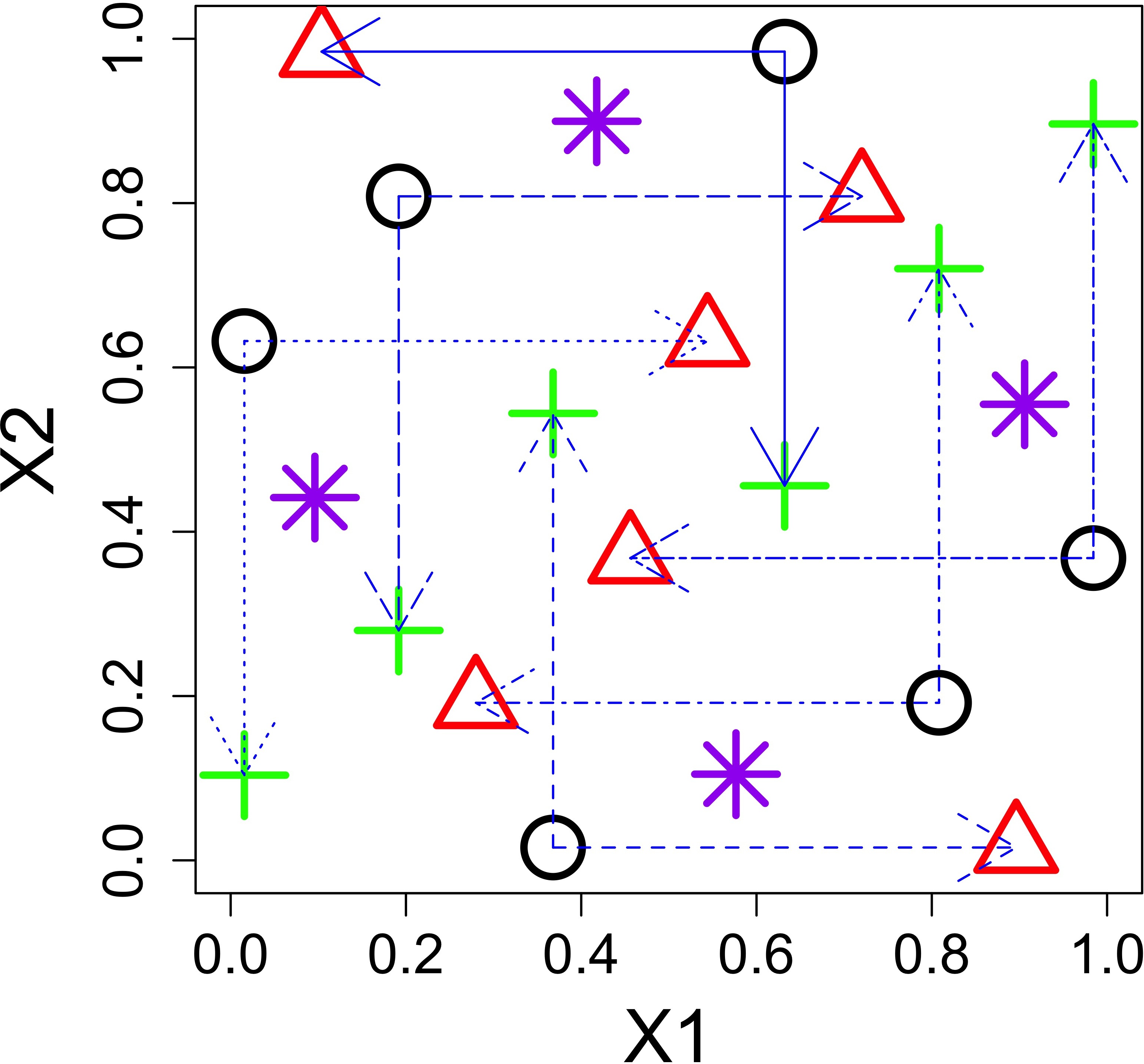}
    \caption{Illustration of an A-SOFT design with $p = 2$, $l = 6$, and $n_{\mathrm{aug}}=4$. The augmented points are shown in purple asterisks.}
    \label{fig:hybrid}
\end{figure}

An illustration of an A-SOFT design with $p=2$, $l=6$, and $n_{\mathrm{aug}}=4$ is shown in Figure~\ref{fig:hybrid}. We next examine the performance of A-SOFT designs in terms of their space-filling properties and surrogate-modeling performance.

\subsection{Space-Filling Properties}
\begin{figure}[!htbp]
    \centering
    \includegraphics[scale=0.73]{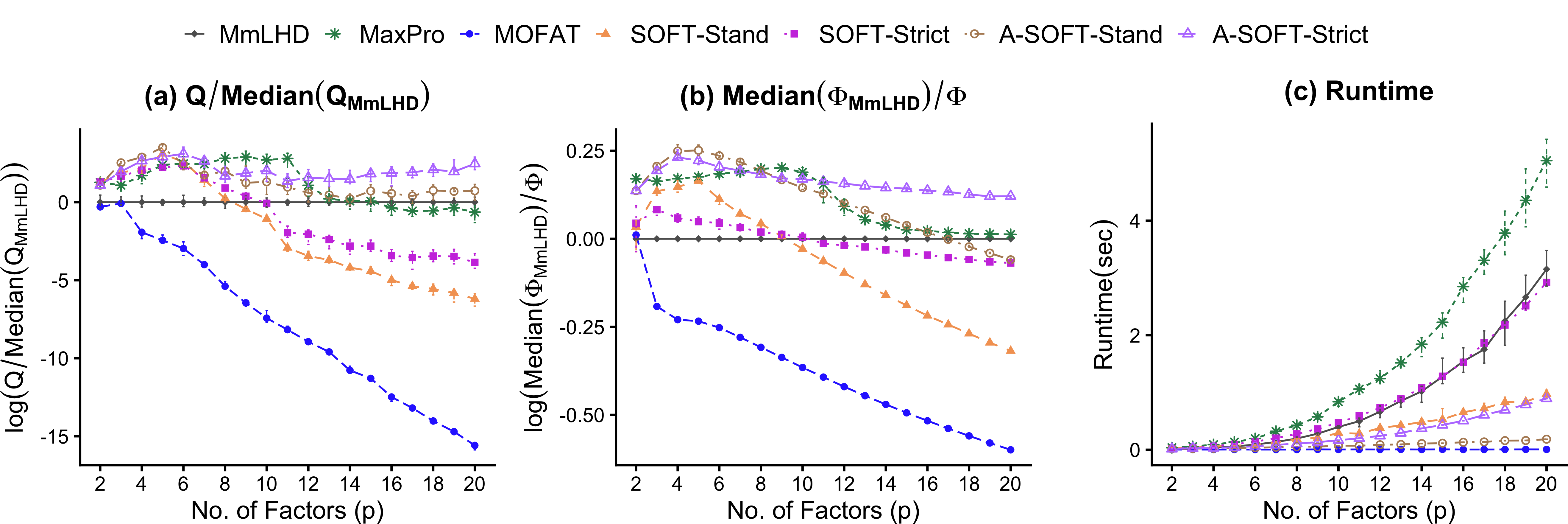}
    \caption{Comparison of the original and augmented designs in $Q$, $\Phi$, and runtime across different numbers of factors.}
    \label{fig:aug_combined_sf}
\end{figure}

To facilitate direct comparison with Section~\ref{subsec:Space-Filling Properties}, we use the same $Q$ and $\Phi$ criteria and experimental settings, with a total run size of $N=6(p+1)$ for $p=2,\ldots,20$. Although the augmented designs are constructed using $\theta=1/m$, where $m$ is the number of levels, all designs are evaluated using $\alpha=1$ and $\theta=1/12$, as in Section~\ref{subsec:Space-Filling Properties}, so that the comparison is based on the same criterion setting. For A-SOFT designs, we use $l=4$ for the SOFT component and allocate the remaining $2(p+1)$ runs to augmentation. As shown in Figure~\ref{fig:aug_combined_sf}, A-SOFT achieves better space-filling performance than the corresponding SOFT designs under both \(Q\) and \(\Phi\), with the improvement becoming more pronounced as \(p\) increases. It also reduces the runtime relative to their unaugmented counterpart because the smaller SOFT component offsets the additional cost of augmentation.

\subsection{Surrogate Modeling}
\begin{figure}[ht]
    \centering
    \includegraphics[scale=0.083]{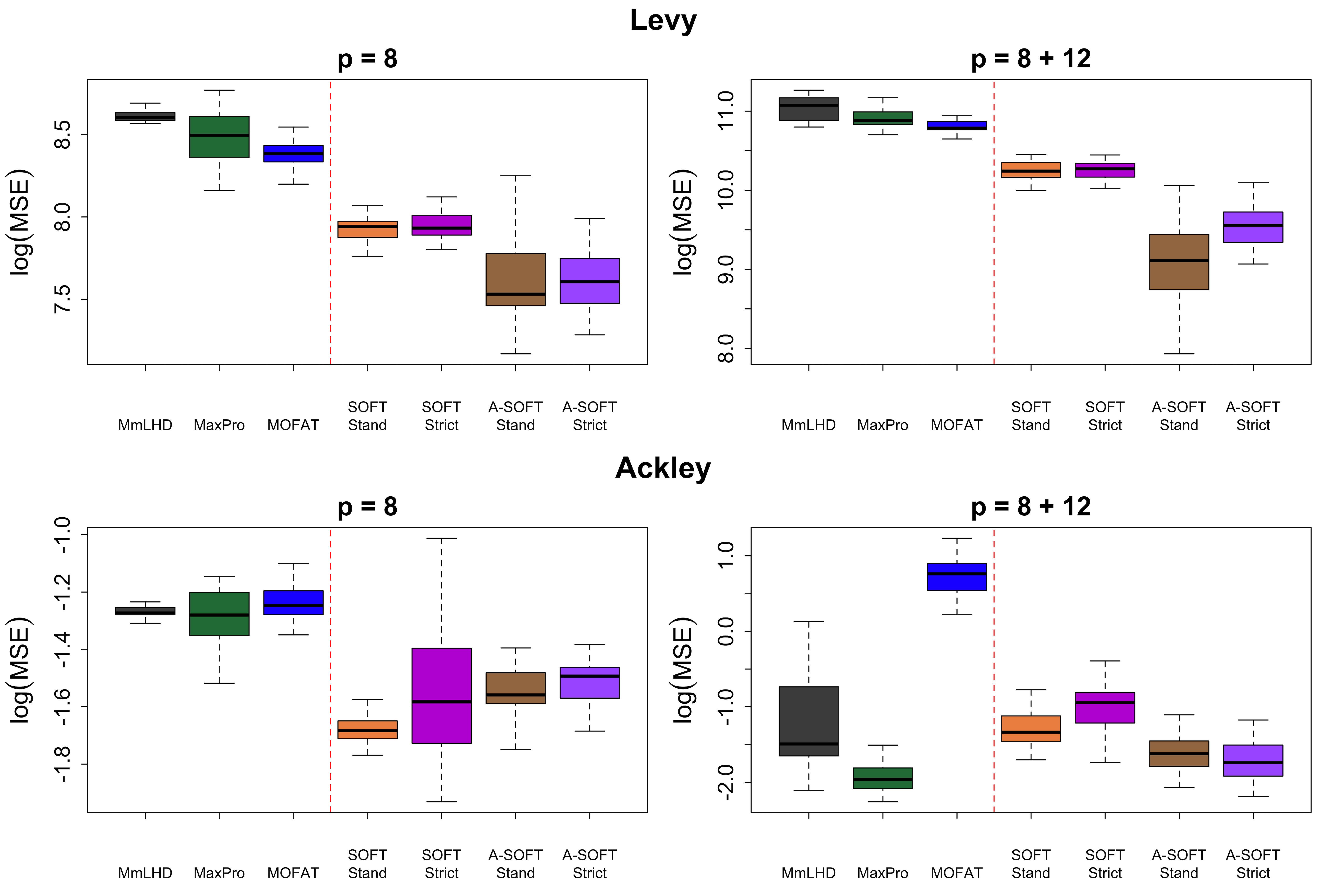}
    \caption{Out-of-sample log MSE of the original and augmented designs for the Levy and Ackley functions under varying input dimensions.}
    \label{fig:aug_combined_mse}
\end{figure}

We revisit the Levy and Ackley functions from Section~\ref{subsec:Surrogate Modeling} using the same experimental settings, with a total run size of \(N=8(p+1)\). For A-SOFT designs, we use \(l=4\) for the SOFT component and allocate the remaining \(4(p+1)\) runs to space-filling augmentation. The results are shown in Figure~\ref{fig:aug_combined_mse}. For Levy function, augmentation substantially improves the predictive performance of SOFT designs for both 8 and 20 dimensional settings. For the Ackley function, augmentation increases the MSE when \(p=8\), but improves predictive performance when \(p=20\). In particular, A-SOFT become competitive with MmLHD and MaxPro in the higher-dimensional setting. Additional simulation results for other benchmark functions are provided in the supplementary materials. 

We also conducted the simulations using a GP model with the Gaussian kernel defined in \eqref{eq:gaussian_kernel}. The results, provided in the supplementary materials, lead to conclusions similar to those obtained using the MIM kernel. This consistency suggests that the improved robustness of SOFT and A-SOFT designs relative to the space-filling designs is not specific to the MIM kernel. Although only two kernels are considered here, the results provide some evidence that this advantage may persist across different choices of GP kernel. 

\begin{figure}[h!]
    \centering
    \includegraphics[scale=0.49]{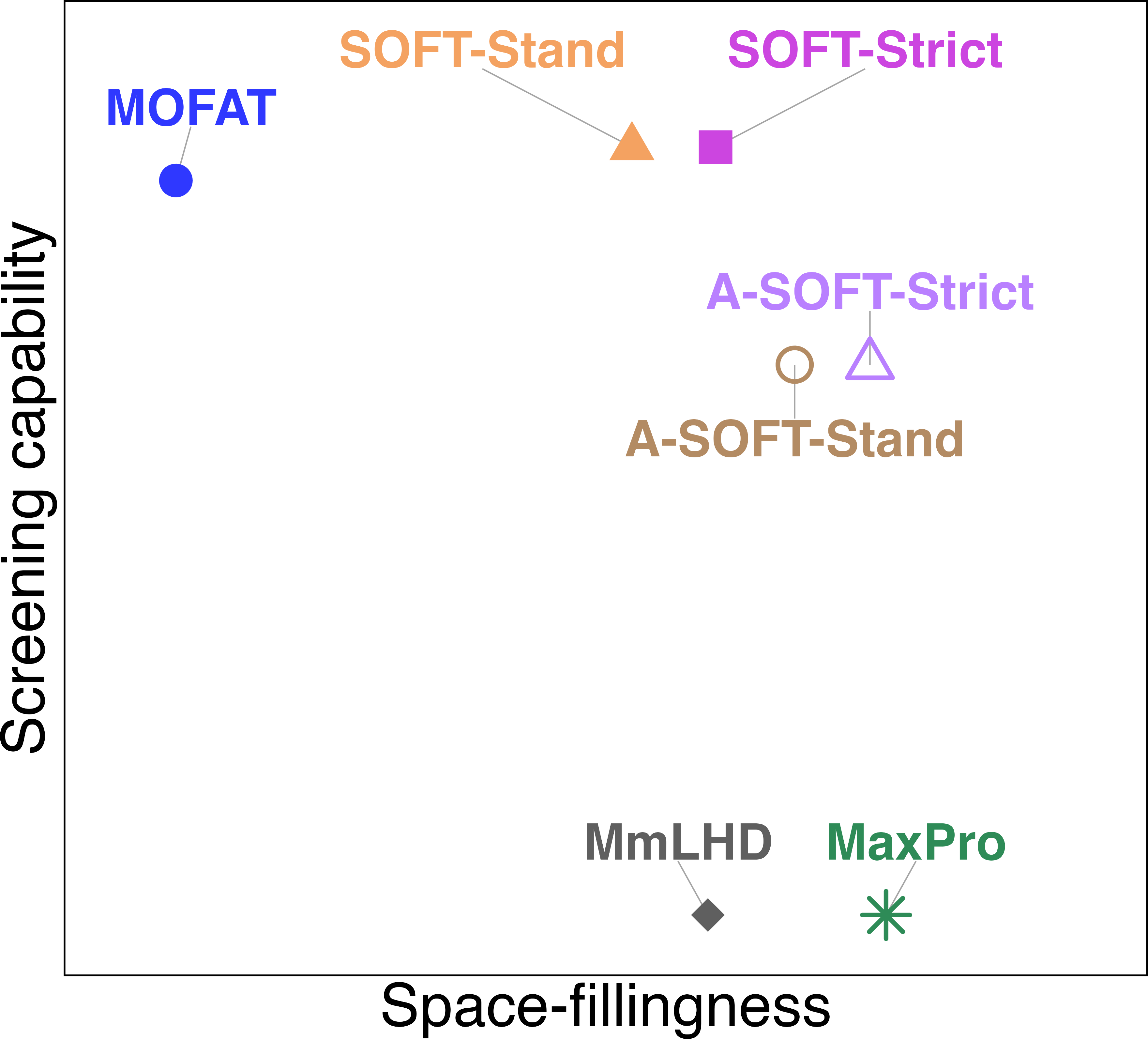}
    \caption{Trade-off between space-fillingness and factor-screening capability for $p=11$ and a total run size of $N=6(p+1)=72$. Space-fillingness is measured by $\log\{Q/\operatorname{Median}(Q_{\mathrm{MmLHD}})\}$ and screening capability is based on the mean perturbation magnitude $\sum_{i=1}^p\sum_{j=1}^{l}|A_{ji}^{(o_i)}-A_{ji}^{(i)}|/p$, which is nonzero only for OFAT-type designs. A-SOFT designs use $l=4$ for the SOFT component and allocate the remaining $2(p+1)=24$ runs for space-filling augmentation.}
    \label{fig:Q_vs_perturbation}
\end{figure}

An advantage of A-SOFT designs over SOFT designs is that there are no restrictions on the number of runs. If $n$ is the budget of the experiment such as $n=10p$ \citep{loeppky2009choosing}, then we can assign about half of the runs to the SOFT component and the remaining to augmentation. This makes A-SOFT designs very flexible. Moreover, A-SOFT designs provide a more robust performance over the other competing designs because they seem to nicely balance the two objectives of screening and space-fillingness, both of which are important for prediction. This trade-off is illustrated in Figure~\ref{fig:Q_vs_perturbation}.

\section{Conclusions}
\label{sec:Conclusion}
Space-filling designs are widely used in deterministic computer experiments. To overcome the challenges arising from effect sparsity, existing methods typically focused on improving the projections of these designs. For example, the popular maximin LHDs that ensure good one-dimensional projections are obtained by finding maximin designs within the class of LHDs. Instead of relying solely on projection properties, a more effective strategy is to identify important factors early, allowing subsequent modeling to focus on the smaller active subspace. However, such a sequential strategy could be impractical when all runs must be launched in a single batch to exploit parallel computing. This motivates the development of a single-shot design that can support both factor screening and surrogate modeling.

Because space-filling designs vary all factors simultaneously, factor screening generally requires a preliminary surrogate modeling \citep{binoisSurveyHighdimensionalGaussian2022} or additional sensitivity-analysis machinery \citep{marrelICSCREAMMethodologyIdentification2022}. In contrast, OFAT designs allow factor importance to be estimated directly from the data without any modeling. However, existing OFAT designs can provide poor coverage of the input space, limiting their usefulness in surrogate modeling. To address this limitation, we proposed SOFT designs, which optimize space-fillingness within the class of OFAT designs. The numerical studies show that SOFT can provide more robust predictive performance, particularly when effect sparsity is present.

Nevertheless, the OFAT structure imposes an inherent limitation. When many inactive factors are present, runs that differ only in those factors collapse to the same locations after projecting onto the active subspace, reducing the number of distinct points available for learning a complex response surface. To alleviate this limitation, we further develop A-SOFT designs, which divide a fixed experimental budget between a SOFT component and space-filling points unrestricted by the OFAT structure. Using fewer OFATs reduces the total perturbation magnitude, which decreases factor-screening capability, but provides greater flexibility for exploring the input space. Simulation results show that A-SOFT designs achieve superior predictive performance compared to both space-filling designs and existing OFAT designs, regardless of the presence of effect sparsity. This indicates that A-SOFT designs are more robust for surrogate modeling than space-filling designs, which is a surprising result. Thus, SOFT designs and their augmented version possess great potential for impact in computer experiments.

The benefits of integrating screening and response surface designs have long been known in the field of physical experiments \citep{cheng2001factor}. Such designs became very popular with the introduction of definitive screening designs \citep{jones2011class}. A recent development in this direction is the orthogonal minimally aliased response surface designs \citep{nunez2020enumeration}. However, we have not come across such integrated designs in computer experiments, where the interest lies in higher-order interactions and nonlinearities. In this regard, SOFT design seems to be the first of its kind. 

\vspace{.25in}
\noindent {\Large\bf Supplementary Materials}

\noindent Proof of Propositions, graphical illustrations, details of test functions, and additional simulations are included in the supplementary file. All scripts required to reproduce all the simulation results are available at \href{https://github.com/weiyangyu/SOFT}{https://github.com/weiyangyu/SOFT}.

\vspace{.25in}
\noindent  {\Large\bf Acknowledgments}

\noindent This research is supported by  a U.S. National Science Foundation grant DMS-2310637.

\bibliographystyle{apalike}
\bibliography{reference}

@article{Sobol_1990aa,
  author    = {I. M. Sobol'},
  title     = {On Sensitivity Estimation for Nonlinear Mathematical Models},
  journal   = {Matematicheskoe Modelirovanie},
  year      = {1990},
  volume    = {2},
  pages     = {112--118},
  note      = {(in Russian)}
}

@article{Sobol_1993bb,
  author  = {I. M. Sobol'},
  title   = {On Sensitivity Estimation for Nonlinear Mathematical Models},
  journal = {Mathematical Modeling and Computational Experiments},
  year    = {1993},
  volume  = {1},
  pages   = {407--414}
}

@article{campolongo2007effective,
  title={An effective screening design for sensitivity analysis of large models},
  author={Campolongo, Francesca and Cariboni, Jessica and Saltelli, Andrea},
  journal={Environmental modelling \& software},
  volume={22},
  number={10},
  pages={1509--1518},
  year={2007},
  publisher={Elsevier}
}

@article{jansen1999analysis,
  title={Analysis of variance designs for model output},
  author={Jansen, Michiel JW},
  journal={Computer Physics Communications},
  volume={117},
  number={1-2},
  pages={35--43},
  year={1999},
  publisher={Elsevier}
}

@article{jones2011class,
  title={A class of three-level designs for definitive screening in the presence of second-order effects},
  author={Jones, Bradley and Nachtsheim, Christopher J},
  journal={Journal of Quality Technology},
  volume={43},
  number={1},
  pages={1--15},
  year={2011},
  publisher={Taylor \& Francis}
}

@article{morris1991factorial,
  title={Factorial sampling plans for preliminary computational experiments},
  author={Morris, Max D},
  journal={Technometrics},
  volume={33},
  number={2},
  pages={161--174},
  year={1991},
  publisher={Taylor \& Francis}
}

@article{xiao2023maximum,
  title={Maximum one-factor-at-a-time designs for screening in computer experiments},
  author={Xiao, Qian and Joseph, V Roshan and Ray, Douglas M},
  journal={Technometrics},
  volume={65},
  number={2},
  pages={220--230},
  year={2023},
  publisher={Taylor \& Francis}
}

@article{homma1996importance,
  title={Importance measures in global sensitivity analysis of nonlinear models},
  author={Homma, Toshimitsu and Saltelli, Andrea},
  journal={Reliability Engineering \& System Safety},
  volume={52},
  number={1},
  pages={1--17},
  year={1996},
  publisher={Elsevier}
}

@article{sobol2001global,
  title={Global sensitivity indices for nonlinear mathematical models and their Monte Carlo estimates},
  author={I. M. Sobol'},
  journal={Mathematics and computers in simulation},
  volume={55},
  number={1-3},
  pages={271--280},
  year={2001},
  publisher={Elsevier}
}

@article{daniel1973one,
  title={One-at-a-time plans},
  author={Daniel, Cuthbert},
  journal={Journal of the American statistical association},
  volume={68},
  number={342},
  pages={353--360},
  year={1973},
  publisher={Taylor \& Francis}
}

@article{mckay1979comparison,
  title={Comparison of Three Methods for Selecting Values of Input Variables in the Analysis of Output from a Computer Code},
  author={McKay, M. D. and Beckman, R. J. and Conover, W. J.},
  journal={Technometrics},
  volume={21},
  number={2},
  pages={239--245},
  year={1979},
  publisher={Taylor \& Francis}
}

@article{zhang2014fractional,
  title={Fractional {B}rownian fields for response surface metamodeling},
  author={Zhang, Ning and Apley, Daniel W},
  journal={Journal of Quality Technology},
  volume={46},
  number={4},
  pages={285--301},
  year={2014},
  publisher={Taylor \& Francis}
}

@article{morris1995exploratory,
  title={Exploratory designs for computational experiments},
  author={Morris, Max D and Mitchell, Toby J},
  journal={Journal of statistical planning and inference},
  volume={43},
  number={3},
  pages={381--402},
  year={1995},
  publisher={Elsevier}
}

@article{johnson1990minimax,
  title={Minimax and maximin distance designs},
  author={Johnson, Mark E and Moore, Leslie M and Ylvisaker, Donald},
  journal={Journal of statistical planning and inference},
  volume={26},
  number={2},
  pages={131--148},
  year={1990},
  publisher={Elsevier}
}

@Manual{Iooss2026sensitivity,
    title = {sensitivity: Global Sensitivity Analysis of Model Outputs and Importance
Measures},
    author = {Bertrand Iooss and Sebastien Da Veiga and Alexandre Janon and Gilles Pujol},
    year = {2026},
    note = {{R} package version 1.31.0},
    url = {https://CRAN.R-project.org/package=sensitivity}
}

@Manual{Xiao2022mofat,
    title = {MOFAT: Maximum One-Factor-at-a-Time Designs},
    author = {Qian Xiao and V. Roshan Joseph},
    year = {2022},
    note = {{R} package version 1.0},
    url = {https://CRAN.R-project.org/package=MOFAT}
}

@Manual{Ba2015SLHD,
    title = {SLHD: Maximin-Distance (Sliced) Latin Hypercube Designs},
    author = {Shan Ba},
    year = {2015},
    note = {{R} package version 2.1-1},
    url = {https://CRAN.R-project.org/package=SLHD}
}

@Manual{Ba2018maxpro,
    title = {Maximum Projection Designs},
    author = {Shan Ba and V. Roshan Joseph},
    year = {2018},
    note = {{R} package version 4.1-2},
    url = {https://cran.r-project.org/package=MaxPro}
}

@Manual{Huang2025rkriging,
    title = {rkriging: Kriging Modeling},
    author = {Chaofan Huang and V. Roshan Joseph},
    year = {2025},
    note = {{R} package version 1.0.2},
    url = {https://cran.r-project.org/package=rkriging}
}

@Manual{Morgan-Wall2025spacefillr,
    title = {spacefillr: Space-Filling Random and Quasi-Random Sequences},
    author = {Tyler Morgan-Wall and Andrew Helmer and Leonhard Grünschloß and Eric Heitz},
    year = {2025},
    note = {{R} package version 0.4.0},
    url = {https://CRAN.R-project.org/package=spacefillr}
}

@book{santner2003design,
  title={The design and analysis of computer experiments},
  author={Santner, Thomas J and Williams, Brian J and Notz, William I and Williams, Brain J},
  volume={},
  year={2003},
  publisher={Springer}
}

@article{saltelli2010avoid,
  title={How to avoid a perfunctory sensitivity analysis},
  author={Saltelli, Andrea and Annoni, Paola},
  journal={Environmental Modelling \& Software},
  volume={25},
  number={12},
  pages={1508--1517},
  year={2010},
  publisher={Elsevier}
}

@book{da2021basics,
  title={Basics and trends in sensitivity analysis: Theory and practice in R},
  author={Da Veiga, S{\'e}bastien and Gamboa, Fabrice and Iooss, Bertrand and Prieur, Cl{\'e}mentine},
  year={2021},
  publisher={SIAM}
}

@article{song2026efficient,
  title={Efficient active learning strategies for computer experiments},
  author={Song, Difan and Joseph, V Roshan},
  journal={Technometrics},
  volume={68},
  number={1},
  pages={65--78},
  year={2026},
  publisher={Taylor \& Francis}
}

@book{saltelli2008global,
  title={Global sensitivity analysis: the primer},
  author={Saltelli, Andrea and Ratto, Marco and Andres, Terry and Campolongo, Francesca and Cariboni, Jessica and Gatelli, Debora and Saisana, Michaela and Tarantola, Stefano},
  year={2008},
  publisher={John Wiley \& Sons}
}

@article{joseph2015maximum,
  title={Maximum projection designs for computer experiments},
  author={Joseph, V Roshan and Gul, Evren and Ba, Shan},
  journal={Biometrika},
  volume={102},
  number={2},
  pages={371--380},
  year={2015},
  publisher={Oxford University Press}
}

@article{krishna2023adaptive,
  title={Adaptive exploration and optimization of materials crystal structures},
  author={Krishna, Arvind and Tran, Huan and Huang, Chaofan and Ramprasad, Rampi and Joseph, V Roshan},
  journal={INFORMS Journal on Data Science},
  volume={3},
  number={1},
  pages={68--83},
  year={2023},
  publisher={INFORMS}
}

@book{gramacy2020surrogates,
  title={Surrogates: Gaussian process modeling, design, and optimization for the applied sciences},
  author={Gramacy, Robert B},
  year={2020},
  publisher={Chapman and Hall/CRC}
}

@article{loeppky2009choosing,
  title={Choosing the sample size of a computer experiment: A practical guide},
  author={Loeppky, Jason L and Sacks, Jerome and Welch, William J},
  journal={Technometrics},
  volume={51},
  number={4},
  pages={366--376},
  year={2009},
  publisher={Taylor \& Francis}
}

@article{sacks1989design,
  title={Design and analysis of computer experiments},
  author={Sacks, Jerome and Welch, William J and Mitchell, Toby J and Wynn, Henry P},
  journal={Statistical science},
  volume={4},
  number={4},
  pages={409--423},
  year={1989},
  publisher={Institute of Mathematical Statistics}
}

@misc{NLopt,
  title = {The {NLopt} nonlinear-optimization package},
  author = {Steven G. Johnson},
  year = {2007},
  howpublished = {\url{https://github.com/stevengj/nlopt}}
}

@book{joseph2026experimental,
  title={Experimental Design for Data Science and Engineering},
  author={Joseph, V Roshan},
  year={2026},
  publisher={Chapman and Hall/CRC}
}

@article{shewry1987maximum,
  title={Maximum entropy sampling},
  author={Shewry, Michael C and Wynn, Henry P},
  journal={Journal of applied statistics},
  volume={14},
  number={2},
  pages={165--170},
  year={1987},
  publisher={Taylor \& Francis}
}

@article{liu2023statistical,
  title={Statistical learning for nonlinear dynamical systems with applications to aircraft-uav collisions},
  author={Liu, Xinchao and Liu, Xiao and Kaman, Tulin and Lu, Xiaohua and Lin, Guang},
  journal={Technometrics},
  volume={65},
  number={4},
  pages={564--578},
  year={2023},
  publisher={Taylor \& Francis}
}

@article{sung2025advancing,
  title={Advancing inverse scattering with surrogate modeling and Bayesian inference for functional inputs},
  author={Sung, Chih-Li and Song, Yao and Hung, Ying},
  journal={SIAM/ASA Journal on Uncertainty Quantification},
  volume={13},
  number={2},
  pages={449--471},
  year={2025},
  publisher={SIAM}
}

@article{cheng2001factor,
  title={Factor screening and response surface exploration},
  author={Cheng, Shao-Wei and Wu, CFJ},
  journal={Statistica Sinica},
  pages={553--580},
  year={2001},
  publisher={JSTOR}
}

@article{nunez2020enumeration,
  title={Enumeration and multicriteria selection of orthogonal minimally aliased response surface designs},
  author={N{\'u}{\~n}ez Ares, Jos{\'e} and Goos, Peter},
  journal={Technometrics},
  volume={62},
  number={1},
  pages={21--36},
  year={2020},
  publisher={Taylor \& Francis}
}

@article{sacks1989designs,
  title={Designs for computer experiments},
  author={Sacks, Jerome and Schiller, Susannah B and Welch, William J},
  journal={Technometrics},
  volume={31},
  number={1},
  pages={41--47},
  year={1989},
  publisher={Taylor \& Francis}
}

@article{oakley2004probabilistic,
  title={Probabilistic sensitivity analysis of complex models: a Bayesian approach},
  author={Oakley, Jeremy E and O'Hagan, Anthony},
  journal={Journal of the Royal Statistical Society Series B: Statistical Methodology},
  volume={66},
  number={3},
  pages={751--769},
  year={2004},
  publisher={Oxford University Press}
}

@article{czitrom1999one,
  title={One-factor-at-a-time versus designed experiments},
  author={Czitrom, Veronica},
  journal={The American Statistician},
  volume={53},
  number={2},
  pages={126--131},
  year={1999},
  publisher={Taylor \& Francis}
}

@article{pujol2009simplex,
  title={Simplex-based screening designs for estimating metamodels},
  author={Pujol, Gilles},
  journal={Reliability Engineering \& System Safety},
  volume={94},
  number={7},
  pages={1156--1160},
  year={2009},
  publisher={Elsevier}
}

@article{santiagoImprovedSensitivityMorris2012,
    title = {Improved sensitivity through {Morris} extension},
    volume = {113},
    issn = {01697439},
    url = {https://linkinghub.elsevier.com/retrieve/pii/S0169743911002103},
    doi = {10.1016/j.chemolab.2011.10.006},
    language = {en},
    urldate = {2026-07-16},
    journal = {Chemometrics and Intelligent Laboratory Systems},
    author = {Santiago, J. and Corre, B. and Claeys-Bruno, M. and Sergent, M.},
    month = apr,
    year = {2012},
    pages = {52--57},
}

@article{saltelli2010variance,
  title={Variance based sensitivity analysis of model output. Design and estimator for the total sensitivity index},
  author={Saltelli, Andrea and Annoni, Paola and Azzini, Ivano and Campolongo, Francesca and Ratto, Marco and Tarantola, Stefano},
  journal={Computer physics communications},
  volume={181},
  number={2},
  pages={259--270},
  year={2010},
  publisher={Elsevier}
}

@article{marrelICSCREAMMethodologyIdentification2022,
    title = {The {ICSCREAM} {Methodology}: {Identification} of {Penalizing} {Configurations} in {Computer} {Experiments} {Using} {Screening} and {Metamodel}—{Applications} in {Thermal} {Hydraulics}},
    volume = {196},
    issn = {0029-5639, 1943-748X},
    shorttitle = {The {ICSCREAM} {Methodology}},
    url = {https://www.tandfonline.com/doi/full/10.1080/00295639.2021.1980362},
    doi = {10.1080/00295639.2021.1980362},
    language = {en},
    number = {3},
    urldate = {2026-07-16},
    journal = {Nuclear Science and Engineering},
    author = {Marrel, A. and Iooss, B. and Chabridon, V.},
    month = mar,
    year = {2022},
    pages = {301--321},
}

@article{binoisSurveyHighdimensionalGaussian2022,
    title = {A {Survey} on {High}-dimensional {Gaussian} {Process} {Modeling} with {Application} to {Bayesian} {Optimization}},
    volume = {2},
    issn = {2688-299X},
    url = {https://dl.acm.org/doi/10.1145/3545611},
    doi = {10.1145/3545611},
    number = {2},
    urldate = {2026-07-15},
    journal = {ACM Trans. Evol. Learn. Optim.},
    author = {Binois, Mickaël and Wycoff, Nathan},
    month = aug,
    year = {2022},
    pages = {8:1--8:26},
}

@book{myers2016response,
  title={Response surface methodology: process and product optimization using designed experiments},
  author={Myers, Raymond H and Montgomery, Douglas C and Anderson-Cook, Christine M},
  year={2016},
  publisher={John Wiley \& Sons}
}

@article{wang2026sfdesign,
  title={{SFD}esign: An {R} Package for Space-Filling Designs.},
  author={Wang, Shangkun and Xie, Weijun and Joseph, V Roshan},
  journal={The R journal},
  volume={18},
  number={2},
  year={2026},
  pages={123--137},
  publisher={R Journal}
}

@book{miller2002subset,
  title={Subset selection in regression, 2nd {E}dition},
  author={Miller, Alan},
  year={2002},
  publisher={CRC Press}
}

\end{document}